\documentclass[useAMS,usenatbib]{mnras}
\usepackage{bm}
\usepackage{graphicx}
\usepackage{booktabs}
\usepackage{siunitx}
\usepackage{enumerate}
\title[Turbulence in the Ionized Gas of the Orion Nebula]{Turbulence
  in the Ionized Gas of the Orion Nebula}
\author[S.~J. Arthur et al.]{S. J. Arthur$^{1}$\thanks{E-mail:
j.arthur@crya.unam.mx}, S.-N. X. Medina$^{2}$, W. J. Henney$^{1}$ \\ 
$^{1}$Instituto de Radioastronom\'{\i}a y Astrof\'{\i}sica, Universidad
Nacional Aut\'onoma de M\'exico, Campus Morelia, 58090 Morelia,
Michoac\'an, M\'exico.\\
$^{2}$Max Planck Instit\"ut f\"ur Radioastronomie, Bonn, Germany.\\
}

\begin{document}
\newcounter{ion}
\newcommand\fakesc[1]{\protect\scalebox{1.0}[0.8]{#1}}
\newcommand\ION[2]{\ensuremath{\mathrm{#1\,\fakesc{#2}}}}
\newcommand\hii{\ion{H}{ii}}
\newcommand\oiii{[\ion{O}{iii}]}
\newcommand\oi{[\ion{O}{i}]}
\newcommand\nii{[\ion{N}{ii}]}
\newcommand\sii{[\ion{S}{ii}]}
\newcommand\siii{[\ion{S}{iii}]}
\newcommand\ha{\ensuremath{\mathrm{H\alpha}}}
\newcommand\kms{\ensuremath{\mathrm{km\ s^{-1}}}}
\newcommand\kmss{\ensuremath{\mathrm{km^2\ s^{-2}}}}
\newcommand\los{\ensuremath{_{\mathrm{los}}}}
\newcommand\pos{\ensuremath{_{\mathrm{pos}}}}
\newcommand\obs{\ensuremath{_{\mathrm{obs}}}}
\newcommand\ins{\ensuremath{_{\mathrm{ins}}}}
\newcommand\rms{\ensuremath{_{\mathrm{rms}}}}
\newcommand\FS{\ensuremath{_{\mathrm{fs}}}}
\newcommand\therm{\ensuremath{_{\mathrm{therm}}}}

\date{}

\pagerange{\pageref{firstpage}--\pageref{lastpage}} \pubyear{2016}

\maketitle

\label{firstpage}
\begin{abstract}
  In order to study the nature, origin, and impact of turbulent
  velocity fluctuations in the ionized gas of the Orion Nebula, we
  apply a variety of statistical techniques to observed velocity
  cubes.  The cubes are derived from high resolving power
  (\(R \approx 40\,000\)) longslit spectroscopy of optical emission
  lines that span a range of ionizations. From Velocity Channel
  Analysis (VCA), we find that the slope of the velocity power
  spectrum is consistent with predictions of Kolmogorov theory between
  scales of 8 and 22 arcsec (0.02 to 0.05~pc).  The outer scale, which
  is the dominant scale of density fluctuations in the nebula,
  approximately coincides with the autocorrelation length of the
  velocity fluctuations that we determine from the second order
  velocity structure function.  We propose that this is the principal
  driving scale of the turbulence, which originates in the
  autocorrelation length of dense cores in the Orion molecular
  filament.  By combining analysis of the non-thermal line widths with
  the systematic trends of velocity centroid versus ionization, we
  find that the global champagne flow and smaller scale turbulence
  each contribute in equal measure to the total velocity dispersion,
  with respective root-mean-square widths of 4--5~km~s$^{-1}$.  The turbulence is
  subsonic and can account for only one half of the derived variance
  in ionized density, with the remaining variance provided by density
  gradients in photoevaporation flows from globules and filaments.
Intercomparison with results from simulations implies that the ionized gas is confined to a thick shell and does not fill the interior of the nebula.

\end{abstract}

\begin{keywords}
hydrodynamics --- \ion{H}{ii} regions --- ISM: individual objects: M~42 --- ISM: kinematics and dynamics --- turbulence
\end{keywords}

\section{Introduction}

Detailed kinematic studies of \ion{H}{ii} regions have shown that the
velocity structure is extremely complex. The spectral lines are
broadened in excess of thermal broadening, suggesting that the gas has
disordered motion along the line of sight. After systematic motions
such as expansion and rotation are accounted for, a random velocity
component still remains. This has been attributed to turbulence in the
photoionized gas. There have been numerous attempts to confirm and
characterize this turbulence in both Galactic and extragalactic
\ion{H}{ii} regions based on statistical methods that examine the
variation of the point-to-point radial velocities with scale (for
example,
\citealp{{1958RvMP...30.1035M},{1985ApJ...288..142R},{1987ApJ...317..686O},
  {1995ApJ...454..316M}, {1997ApJ...487..163M},
  {2011MNRAS.413..721L}}, and references cited by these papers).

Structure functions of velocity centroids have become a standard
statistical tool since \citet{1951ZA.....30...17V} investigated the
projection of a three-dimensional correlation function onto the plane
of the sky. They measure the variation of the mean velocity integrated
along the line of sight as a function of the plane-of-sky separation.
However, the construction of structure functions over a wide range of
scales requires high velocity resolution data with extensive spatial
coverage. This can be realized by multiple high-velocity-resolution
longslit spectroscopic observations of an optically thin emission line
at a variety of positions across an \ion{H}{ii} region
\citep{{1987ApJ...317..676O},{1987ApJ...315L..55C},{1993ApJ...409..262W}},
or by Fabry-Perot interferometry
\citep{{1985ApJ...288..142R},{1995ApJ...454..316M},{2011MNRAS.413..705L}},
which has much lower velocity resolution.

The observational requirements of high velocity resolution coupled
with high spatial resolution and coverage mean that M42, the Orion
Nebula, is ideally suited for this type of study. It is the closest
and brightest \ion{H}{ii} region ($\sim 440$~pc, see e.g.,
\citealp{{2008AJ....136.1566O}}) and its geometry, main kinematic
features and stellar population are well known (see O'Dell
\citealp{{2001ARAA..39...99O}} for a review). The inner part of the
nebula measures some $3^{\prime}\times 5^{\prime}$ ($0.38 \times
0.63$~pc at the quoted distance) and is bright in optical emission
lines such as [\ion{S}{ii}]\,$\lambda$6716,6731,
[\ion{N}{ii}]\,$\lambda$6583, H$\alpha$ and
[\ion{O}{iii}]\,$\lambda$5007. The principal ionizing star is
$\theta^1$Ori~C (spectral type $\sim$O7), which also possesses a fast
stellar wind \citep{2009AJ...137..367O}, and there are also 3 B-type
stars. These 4 early type stars are the Trapezium cluster. M42 is a
site of ongoing star formation on the nearside of the Orion Molecular
Cloud (OMC-1) and there is a large population of young stars, some of
which are the sources of stellar jets and Herbig-Haro objects and some
have associated protoplanetary discs (\textit{proplyds},
\citealp{1993ApJ...410..696O}). M42 is an example of a blister
\ion{H}{ii} region
\citep{{1973ApJ...183..863Z},{1974PASP...86..616B},{1979AA....71...59T},{2009AJ...137..367O}}
and the photoionized gas is streaming away from the background
molecular cloud with blueshifted velocities of order 10~km\,s$^{-1}$.

Early attempts at characterizing the turbulence in M42 using the
[\ion{O}{iii}]\,$\lambda$5007 emission line were inconclusive and
suffered from poor spatial coverage
\citep{{1951ZA.....30...17V},{1958RvMP...30.1035M}}.  Detailed
kinematic studies of the [\ion{O}{iii}]\,$\lambda$5007 emission line
\citep{1988ApJS...67...93C} and the [\ion{S}{iii}]\,$\lambda$6312 emission
line \citep{1993ApJ...409..262W} in the inner region of the Orion
Nebula, which decomposed the line profiles into individual components,
found evidence for the presence of turbulence. They obtained
statistical correlations in the structure functions of the principal
line component at spatial scales $< 15^{\prime\prime}$. However, the
power-law indices do not agree with those expected from the standard
Kolmogorov theory for isothermal, homogeneous turbulence nor with
those predicted by \citet{1951ZA.....30...17V}, taking into account
projection smoothing. \citet{1992ApJ...387..229O} carried out a
similar study for [\ion{O}{i}]\,$\lambda$6300. Although their structure
functions appear to agree with Kolmogorov theory for a wide range of
spatial scales, there is a problem with the line widths, which are
much larger than they should be for the calculated structure function.
\citet{1992ApJ...387..229O} conclude that the [\ion{O}{i}] emission comes from
the vicinity of the highly irregular ionization front, and that the
large line widths are due to the acceleration of the gas away from the
corrugated surface.

Other statistical techniques have been used to study turbulence in the
interstellar medium. In particular, velocity channel analysis
\citep{2000ApJ...537..720L} is optimally applied to spatially resolved
PPV data from optically thin emission lines and can be used when the
turbulence is supersonic, unlike the velocity centroid statistics,
which are best employed when the turbulence is subsonic or at most
mildly supersonic \citep{2004ASSL..319..601L}. It has been applied to
the HI gas in the Galaxy \citep{2000ApJ...537..720L} and in the SMC
\citep{2001ApJ...551L..53S}. There are no previous VCA studies of the
ionized gas in the Orion Nebula (or any other ionized gas VCA studies
of which we are aware).

In a previous paper (\citealp{2014MNRAS.445.1797M}, hereafter Paper~I)
we investigated the scale dependence of fluctuations inside a
realistic model of an evolving turbulent \ion{H}{ii} region. We
calculated structure functions of velocity centroid maps for a variety
of simulated emission lines but found that these were not a reliable
way to recover the intrinsic 3D power spectrum of the velocity
fluctuations of our numerical simulation. We found that the velocity
channel analysis technique was a more promising approach, despite
being intrinsically limited by thermal broadening. Using this method,
we successfully recovered the logarithmic slope of the underlying
velocity power spectrum to a precision of \(\pm 0.1\) from simulated
high resolution optical emission-line spectroscopy. Furthermore, our
simulations revealed that multiple scales of energy injection due to
champagne flows and the photoionization of clumps and filaments result
in a flatter spectrum of fluctuations than would be expected from
Kolmogorov theory of top-down turbulence driven at the largest scales.

In this paper, we apply both velocity channel analysis of
position-velocity arrays and structure functions of velocity centroid
maps to high spatial and velocity resolution observations of low- and
high-ionization emission lines of the central $3^{\prime}\times
5^{\prime}$ region of the Orion Nebula. These observations were 
presented in the form of an atlas of emission lines by
\cite{2007AJ....133..952G} and \citet{2008RMxAA..44..181G}. The
longslit observations cover the full area at $2^{\prime\prime}$
separation and allow maps to be made of the velocity moments. These
maps are used to calculate the structure functions. In addition, we
have the position-velocity arrays of the individual longslit
observations, with which we can perform the velocity channel
analysis. We investigate whether the structure function method and the
velocity channel analysis give the same information about the nature
of the turbulence in the photoionized gas of the Orion Nebula. We also
compare the results to our work in Paper I to assess to what extent
numerical simulations can produce the detailed kinematics of
\ion{H}{ii} regions.

The paper is structured as follows: in Section~\ref{sec:methods} we
describe the observational data and the statistical methods used in
this work. Section~\ref{sec:results} presents the results of the
velocity channel analysis and the structure function calculations. In
Section~\ref{sec:discuss} we discuss our findings with respect to
previous observational work and with regard to the numerical
simulations of Paper I. We combine these results with complementary
information derived from studies of linewidths and surface brightness
fluctuations to form a picture of the causal links between physical
processes and velocity and brightness fluctuations in the Orion Nebula.
The conclusions are summarized in \S~\ref{sec:summary-conclusions}.

\section[]{Methods}
\label{sec:methods}
\begin{figure*}
\centering
\includegraphics[width=0.9\textwidth]{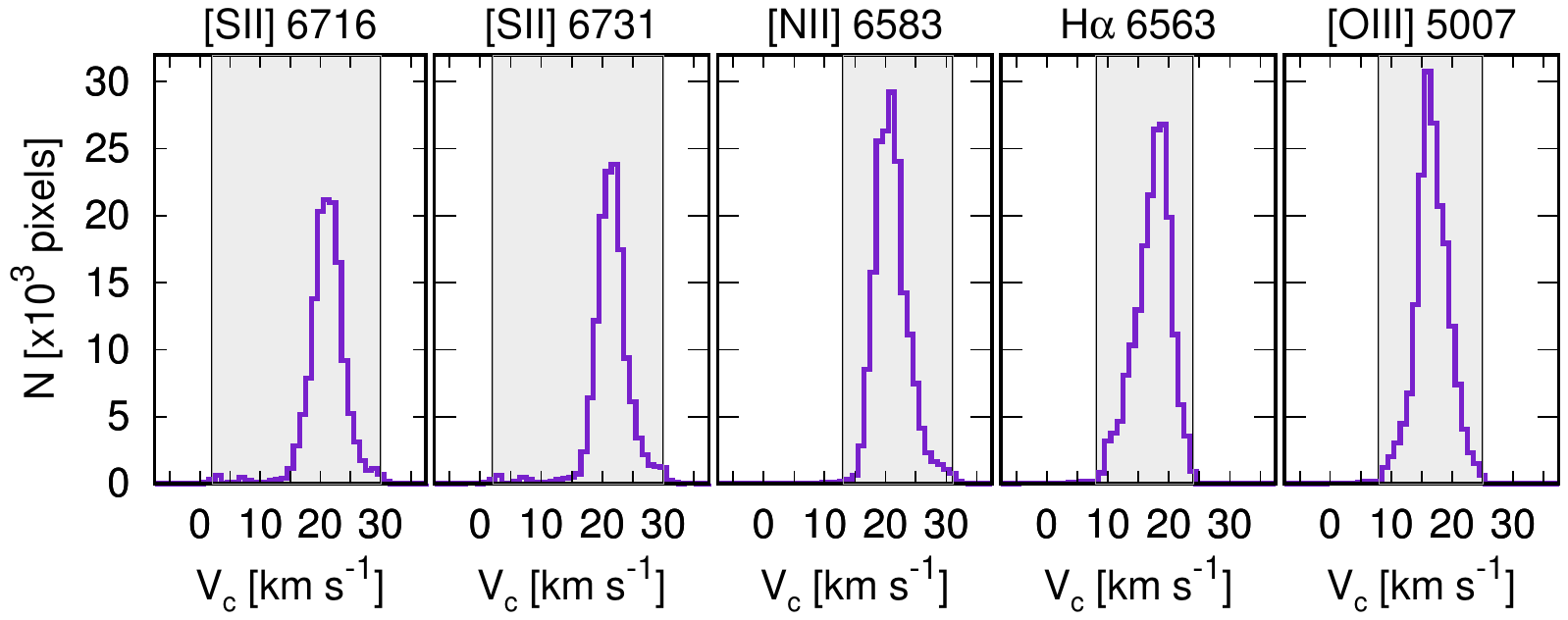}\\
\includegraphics[width=0.9\textwidth]{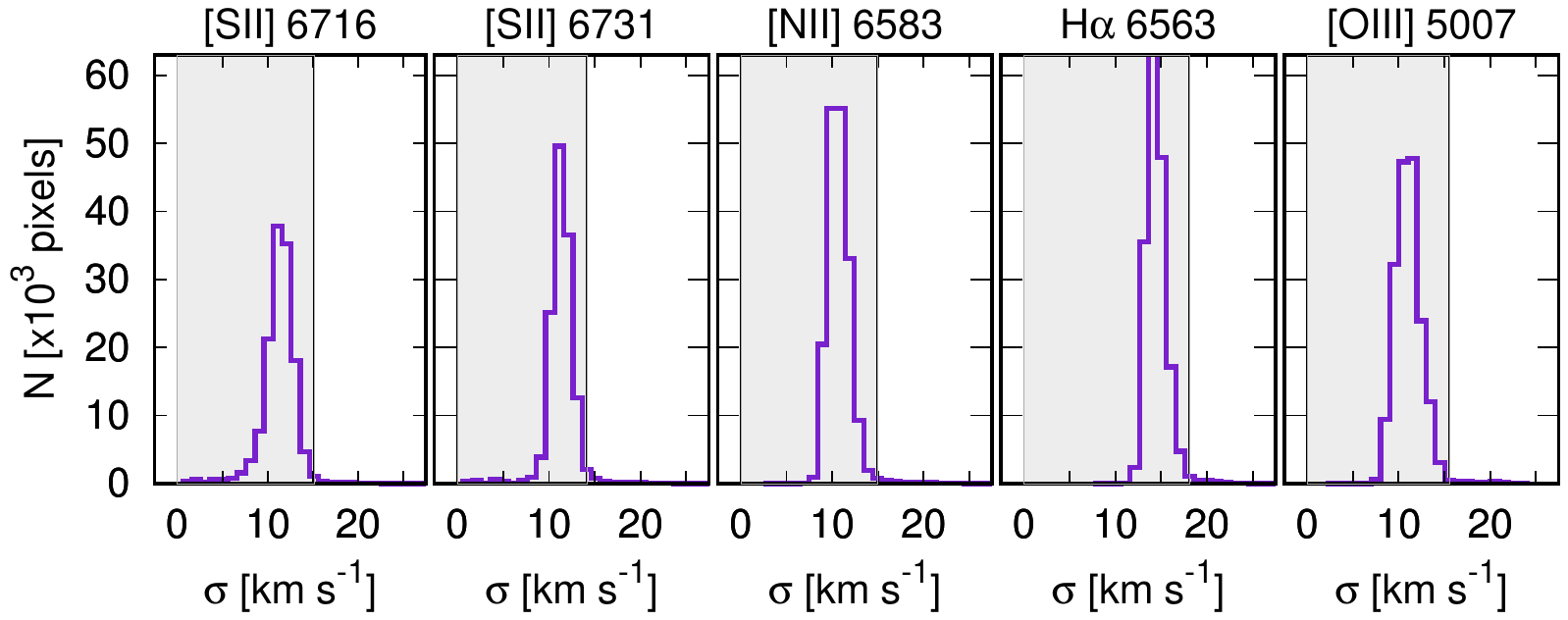}
\caption{Probability density functions (PDF) of the [\ion{S}{ii}]\,$\lambda$6716\,\AA\ and
  [\ion{S}{ii}]\,$\lambda$6731\,\AA\, [\ion{N}{ii}]\,$\lambda$6584\,\AA, H$\alpha$ and
  [\ion{O}{iii}]\,$\lambda$5007\,\AA, emission line velocity centroids (top
  panel) and velocity dispersions (bottom panel). The shaded grey areas
indicate the selected ranges of the centroid velocities and velocity
dispersions used to create masks. The velocity dispersions have not
been corrected for thermal broadening nor for the instrumental width.}
\label{fig:pdf}
\end{figure*}
\subsection{Observational Data}
The spectral line data was presented in the emission line atlas of
\citet{2008RMxAA..44..181G}. In this paper, we use the H$\alpha$, [\ion{S}{ii}] 6716,
6731~\AA, 
[\ion{N}{ii}]~6583~\AA\ and [\ion{O}{iii}]~5005~\AA\ data obtained with the echelle
spectrograph attached to the 4~m telescope at Kitt Peak National
Observatory (KPNO; for
observational details see \citealp{2004AJ....127.3456D}), with
supplementary [\ion{S}{ii}] data observed with the Manchester Echelle Spectrometer attached
to the 2.1~m telescope at the Observatorio Astron\'omico Nacional at
San Pedro M\'artir (OAN-SPM), Mexico. The data
covers the $3^{\prime}\times 5^{\prime}$ central (Huygens) region of the Orion
Nebula\footnote{ The total observed area is from
  $5^{\mathrm{h}}35^{\mathrm{m}}10.4^{\mathrm{s}}$ to
  $5^{\mathrm{h}}35^{\mathrm{m}}23.2^{\mathrm{s}}$ in right ascension and
  from  $-5^{\circ} 21^{\prime} 36^{\prime\prime}$  to  $-5^{\circ}
  26^{\prime}10^{\prime\prime}$ in declination. The position of
  $\theta^1$Ori~C is R.A.~$5^{\mathrm{h}} 35^{\mathrm{m}}
  16.5^{\mathrm{s}}$ Dec. $-5^{\circ} 23^{\prime} 23^{\prime\prime}$
  (J2000)}   and the H$\alpha$, [\ion{N}{ii}] and [\ion{O}{iii}] data consists of  96 approximately $300^{\prime\prime}$ 
North-South orientated slits at $2^{\prime\prime}$ intervals, where
the slit width corresponds to $0.8^{\prime\prime}$. The velocity
resolution of these observations is 8~km~s$^{-1}$. The [\ion{S}{ii}] 6716,
6731~\AA\ dataset has a total of 92 North-South pointings covering the
same region, consisting of 37 positions observed at KPNO and 55
positions observed at OAN-SPM.
The KPNO [\ion{S}{ii}] data consists of two disjoint
regions, one in the east and one in the west of the nebula,
while the OAN-SPM data consists of 35 pointings with a slit width
corresponding to $2^{\prime\prime}$ and 12~km~s$^{-1}$ velocity
resolution, and 20 pointings with a $0.9^{\prime\prime}$ slit width and
6~km~s$^{-1}$ velocity resolution. The slit length at OAN-SPM is
$312^{\prime\prime}$. 

In the present paper, we use the calibrated position-velocity (PV)
arrays obtained by \citet{2008RMxAA..44..181G} from each individual longslit spectrum for the
velocity channel analysis of each emission line. For the second-order
structure functions, we work with the velocity moment maps,
particularly the mean velocity map (i.e., map of velocity centroids)
reconstructed from the PV arrays by \citet{2008RMxAA..44..181G}.

\subsection{Statistical Methods}

 To analyze the velocity fluctuations due to turbulence in the ionized gas, we
 make use of two statistical tools: velocity channel analysis (VCA) of the PV
 data arrays and second-order structure functions
 of the velocity centroid maps.

\subsubsection{Velocity Channel Analysis}
\label{subsec:MVCA}
VCA consists of taking the spatial power spectrum
of the emission intensity (or brightness) in velocity channels of spectroscopic PV data. 
First, the PV
data is binned into velocity channels of width $\delta v$. The relative contribution of
velocity fluctuations to fluctuations in the total intensity decreases
as the width of the velocity slices increases, because thicker
velocity slices average out the contributions of many velocity
fluctuations. The very thickest velocity slice gives information only on
the density spectral index. A velocity slice has width $\delta v
= (v_\mathrm{max} - v_\mathrm{min})/N$, where $N$ is the number of
channels and for the spectra used in this work $v_\mathrm{max} =
70$~km\,s$^{-1}$ and $v_\mathrm{min} = -40$~km\,s$^{-1}$. 

\defcitealias{2014MNRAS.445.1797M}{Paper I}
The thickest velocity channel ($N = 1$)
corresponds to the total velocity integrated emission-line surface brightness at each pixel along the slit. The
thinnest velocity slices for the present work are chosen to have $\delta v =
4$~km\,s$^{-1}$, to give a good sampling of the spectrograph velocity
resolution (FWHM) of 6--8~km\,s$^{-1}$ for all the emission lines.
To use thinner velocity slices would be to amplify the noise (see
\citealp{2014MNRAS.445.1797M}, hereafter Paper~I). 

The power spectra for each velocity slice of each PV array 
for a given emission line are summed and normalized by the
total power. Since each spectroscopic observation corresponds
to a finite slit length, we use a Welch window function to reduce the
edge effects in the estimated power spectrum \citep{1992nrfa.book.....P}
before applying the fast Fourier transform (FFT).

\subsubsection{Second-order Structure Functions}
\label{sssec:SOSF}
The second-order structure function of the velocity centroids is (see \citetalias{2014MNRAS.445.1797M})
\begin{equation}
  S_2(\boldsymbol{l}) = \frac{\sum_\mathrm{pairs}[V_c(\mathbf{r}) - V_c(\mathbf{r} +
    \boldsymbol{l})]^2}{\sigma_\mathrm{vc}^2 N(\boldsymbol{l})} \ .
\label{eq:sf}
\end{equation}
In this definition, $\mathbf{r}$ is the
two-dimensional position vector in the plane of the sky, while
$\boldsymbol{l}$ is the separation vector. The normalization is by the
number of pairs of points at each separation,  $N(\boldsymbol{l})$,
and the variance of centroid velocity fluctuations, $\sigma^2_\mathrm{vc}$, defined by
\begin{equation}
\sigma^2_\mathrm{vc} \equiv \frac{\sum_\mathrm{pixels} [V_c(\mathbf{r}) - \langle V_c \rangle
  ]^2}{N} \ .
\label{eq:variance}
\end{equation}
Here, $\langle V_c \rangle$ is the mean centroid velocity 
\begin{equation}
  \langle V_c \rangle \equiv \frac{\sum_\mathrm{pixels} V_c(\mathbf{r})}{N} \ . 
\label{eq:mean}
\end{equation}
The summation in Equation~\ref{eq:sf} is over all data pairs for each
separation, $N(\boldsymbol{l})$, while the summations in the centroid
variation and mean (Eqs.~\ref{eq:variance} and \ref{eq:mean}) are over the total number of array elements, i.e.,
valid pixels in the $(x,y)$-plane.

We also define an intensity weighted structure function by 
\begin{equation}
  S_2(\boldsymbol{l}) = \frac{\sum_\mathrm{pairs}[V_c(\mathbf{r}) - V_c(\mathbf{r} +
    \boldsymbol{l})]^2 I(\mathbf{r}) I(\mathbf{r} + \boldsymbol{l}) }{\sigma_\mathrm{vc}^2 W(\boldsymbol{l})} \ ,
\label{eq:sfw}
\end{equation}
where $W(\boldsymbol{l}) = \sum_\mathrm{pairs} I(\mathbf{r})
I(\mathbf{r} + \boldsymbol{l}) $ is the sum of the weights for each
separation and $I(\mathbf{r})$, $I(\mathbf{r} + \boldsymbol{l})$ are the
weights (i.e., intensities) of each pair of pixels. This form of the structure
function obviously favours bright structures and reduces the
contribution of fainter regions. This is one way to reduce the
contribution of noise to the structure function.

The structure function is affected by small-scale, high velocity features such as jets and Herbig-Haro
 objects and we aim to eliminate these pixels
since they are not associated with the underlying turbulence.
 The first step is to examine the probability density function (pdf)
 of the velocity centroid maps and the corresponding velocity
 dispersion maps of the H$\alpha$\,$\lambda$6583, [\ion{N}{ii}]\,$\lambda$6853,
 [\ion{O}{iii}]\,$\lambda$5007, [\ion{S}{ii}]\,$\lambda$6716 and [\ion{S}{ii}]\,$\lambda$6731
 lines from the emission line atlas published by
 \citet{2008RMxAA..44..181G}. 
 In practice, a 2\% threshold was
 uniformly applied to the velocity centroid pdf binned at
 1~km~s$^{-1}$ resolution. The same threshold was also applied to one
 side of the velocity dispersion pdf binned at 1~km~s$^{-1}$ resolution
 to eliminate the small number of pixels with anomolously high
 values. Figure~\ref{fig:pdf} shows the initial pdfs of the centroid
 velocities and the velocity dispersions of the 5 emission lines
 studied in this paper. Note that the velocity dispersions have not
 been corrected for the instrumental width (around 3~km~s$^{-1}$) nor
 thermal broadening. Thermal broadening most affects the H$\alpha$
 line, and is responsible for the peak of the velocity dispersion
 distribution occurring at $\sigma = 15$~km~s$^{-1}$, instead of the
 lower value of $\sigma = 11$~km~s$^{-1}$ seen for the other emission
 lines. Figure~\ref{fig:pdf} also highlights in grey the range of
 values used to create masks of ``good'' pixels for the subsequent
 structure function calculations. Pixels having $V_\mathrm{c}$ or
 $\sigma$ values outside these ranges are excluded from the structure
 function analysis.
\begin{figure*}
  \centering
\includegraphics[height=0.7\textheight]{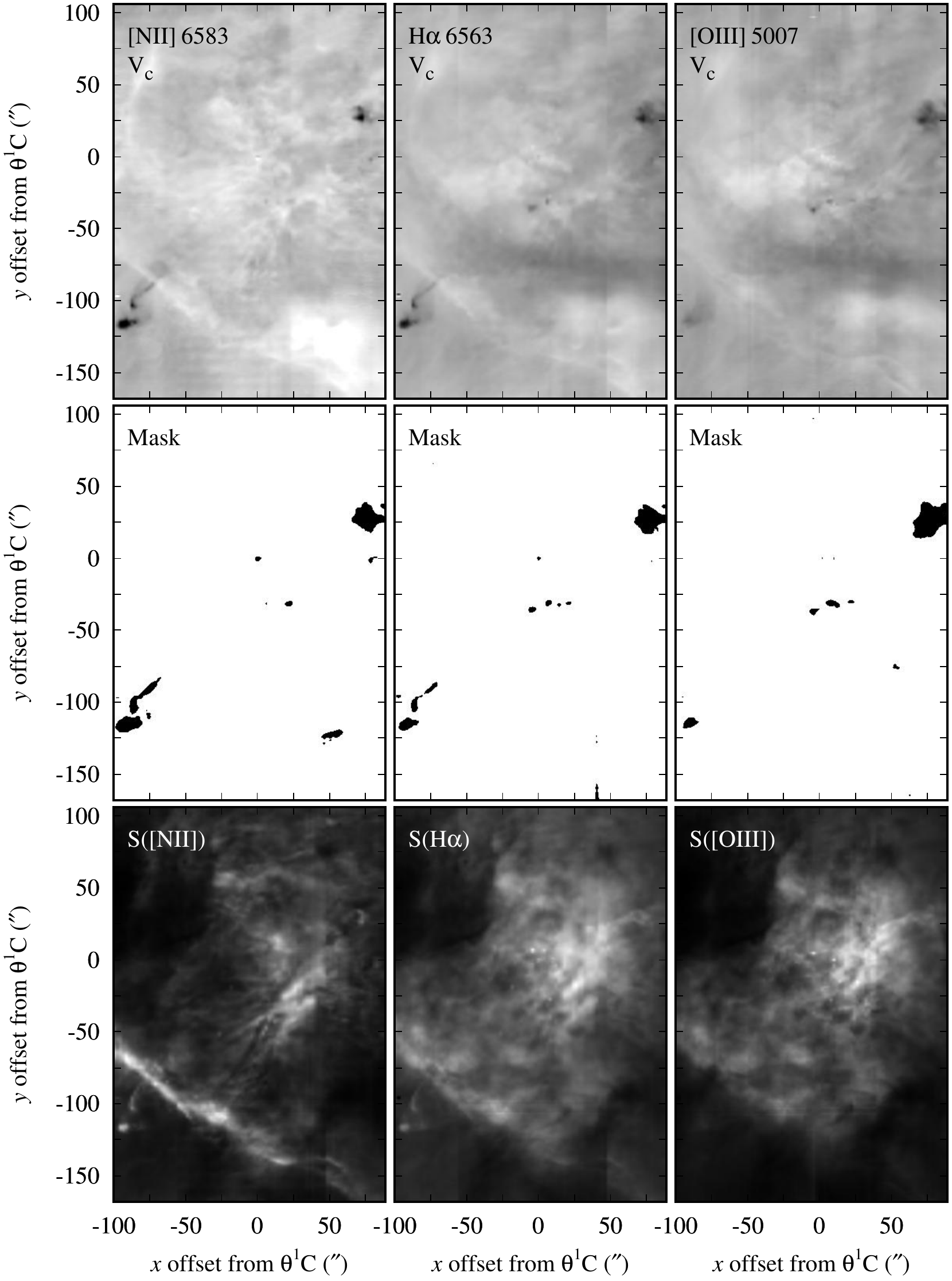}
\caption{Two-dimensional interpolated maps of velocity centroids of
  the (left to right) [\ion{N}{ii}]\,$\lambda$6584\,\AA, H$\alpha$ and
  [\ion{O}{iii}]\,$\lambda$5007\,\AA\ emission lines. The upper panels show the
  initial 2D interpolated velocity centroid maps with a linear
  greyscale between $-40$ and 70~km~s$^{-1}$. The centre panels
  show the masks obtained by eliminating pixels with extreme values of
  the centroid velocity ($V_\mathrm{c}$) or velocity dispersion
  ($\sigma$), as indicated in Fig.~\ref{fig:pdf}. The lower panels show
  the integrated intensity (surface brightness) with a linear
  greyscale. In these images, North is up and East is to the left and
  the $x$ and $y$ scales give the offset in arcsec with respect to the
  position of the main ionizing star of the Orion Nebula, $\Theta^1$Ori~C.
}
\label{fig:maskHNO}
\end{figure*}
\begin{figure*}
  \centering
\includegraphics[height=0.7\textheight]{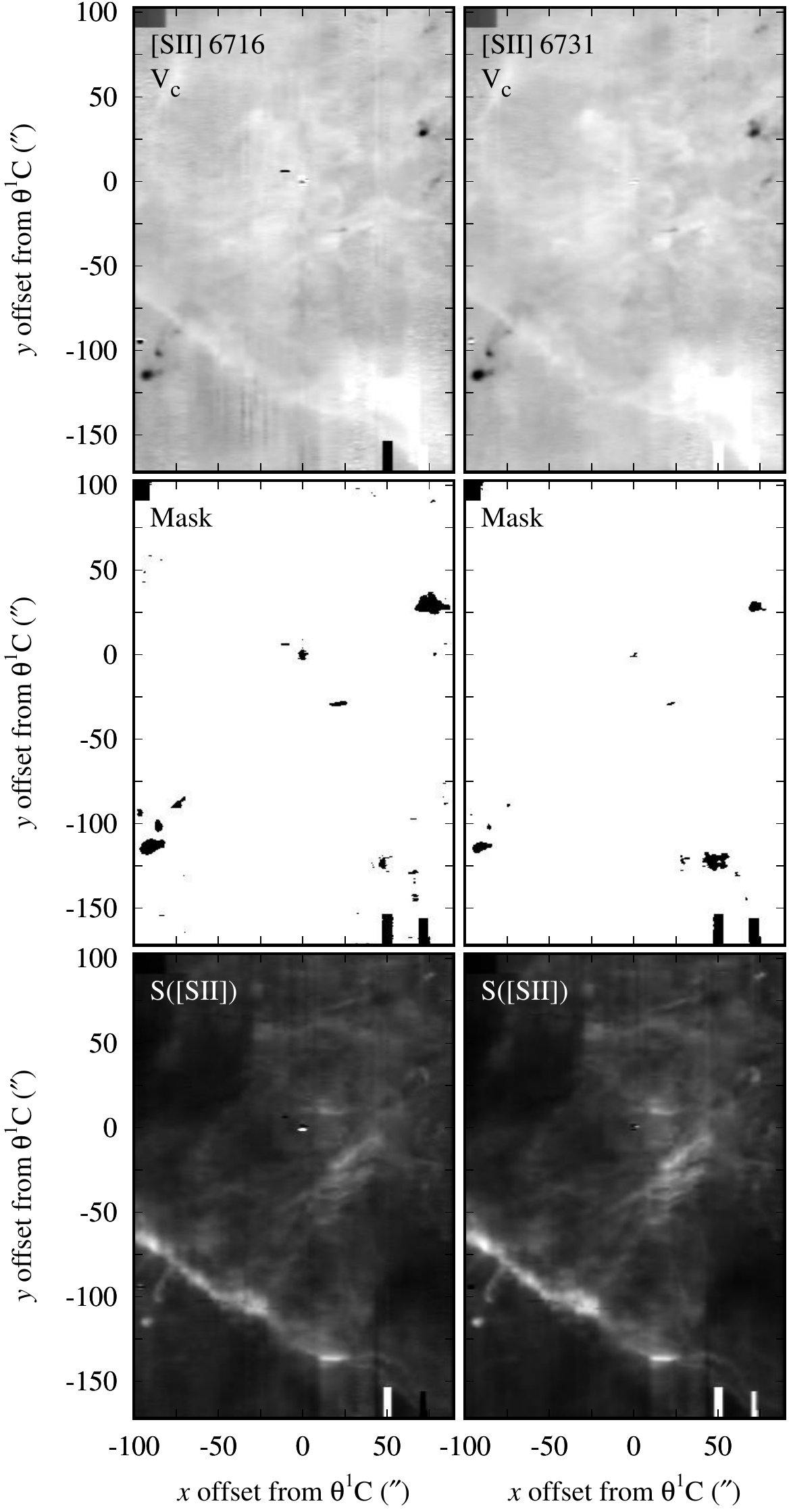}
\caption{Same as Fig.~\protect\ref{fig:maskHNO} but for the [\ion{S}{ii}]\,$\lambda$6716\,\AA\ and
  [\ion{S}{ii}]\,$\lambda$6731\,\AA\ emission lines.}
\label{fig:masksii}
\end{figure*}

Figures~\ref{fig:maskHNO} and \ref{fig:masksii} show the 2D
interpolated velocity centroid maps together with the masks
constructed from pixels with extreme values of the centroid velocity
or velocity dispersion identified through consideration of the
pdfs. It can be seen that these pixels correspond to high velocity
features, such as HH~201, 202, 203, 204, and 529, and also to
numerical artefacts introduced during the original construction of the
maps from the individual slits.

\subsection{Application to the Orion Nebula}

Although the standard theory assumes that the turbulent velocity
fluctuations are homogeneous and isotropic, the real
Orion Nebula is a complex, inhomogeneous object with ill-defined
boundaries. We expect variation in the power-law indices obtained from
both the velocity channel analysis and the second-order structure
functions depending on position in the nebula due to large-scale
inhomogeneities. We estimate the confidence bounds on the power-law indices by
resampling. 

For the velocity channel analysis, this takes the form of a bootstrap
Monte Carlo method, that is, resampling with replacement of the set of
PV arrays.  The power-law indices of the power spectra resulting from
10 different resamplings of the set of PV arrays (96 arrays in the
case of the [\ion{N}{ii}], H$\alpha$ and [\ion{O}{iii}] emission lines and 20 arrays
for [\ion{S}{ii}] --- see Fig.~\ref{fig:siislits} for the positions of the
high-resolution [\ion{S}{ii}] slits) are obtained using a least-squares fitting procedure. The
variation of these sample power-law indices provides an estimate of
the confidence bounds for the power-law index across the Orion Nebula.

Furthermore, our longslit spectra are all oriented North-South and so in
order to check whether this has any effect on the power-law index, we
analyze a supplementary PV dataset of [\ion{O}{iii}]\,$\lambda$5007 observations
from 18 slit positions perpendicular to the main data set (i.e.,
oriented East-West). The slit positions and orientations are indicated
in Figure~\ref{fig:ohoriz} \citep{2015AJ....150..108O}.

\begin{figure}
 \centering
\includegraphics[width=\columnwidth]{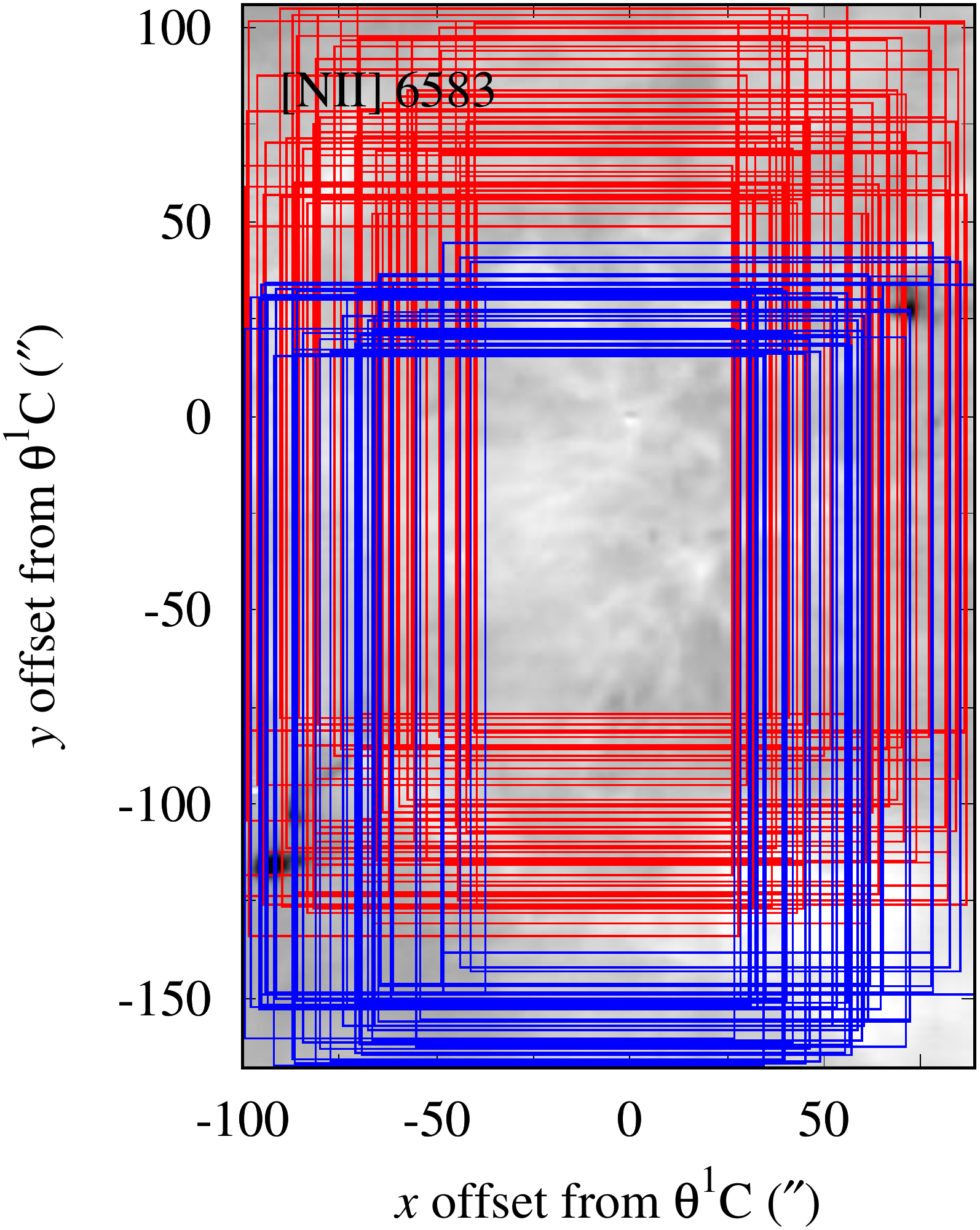}
 \caption{Distribution of randomly selected rectangular frames for the
   structure function calculations for [\ion{N}{ii}]. The [\ion{N}{ii}] frames show the spatial separation of the
   two distinct structure function populations (depicted with red and
   blue rectangles).}
 \label{fig:frames}
 \end{figure}

We estimate the effects of large-scale spatial inhomogeneity
on the second-order structure function
by evaluating $S_2(\boldsymbol{l})$ (see Eqs.~\ref{eq:sf} and \ref{eq:sfw}) for 100 distinct,
randomly selected rectangular frames within each 2D map. Each frame
has two-thirds the $x$ and $y$ dimension of the initial map (see Fig.~\ref{fig:frames}). 
The variation in power-law indices of
the smaller maps give us an estimate of 
the spread of power-law behaviour due to location. 

\section[]{Results}
\label{sec:results}
We present the results of the velocity channel analysis in the form of
power spectra plotted against wavenumber $k$ in units of number of
waves per arcsecond and also number of waves per parsec. The conversion to
parsec assumes a distance of 440~pc to the Orion Nebula \citep{2008AJ....136.1566O}. The
second-order structure
functions are plotted against separation scale in both arcseconds
and parsec.
\subsection{Velocity Channel Analysis}
\label{subsec:vca}
\begin{figure*}
\centering
\includegraphics[width=0.75\textwidth]{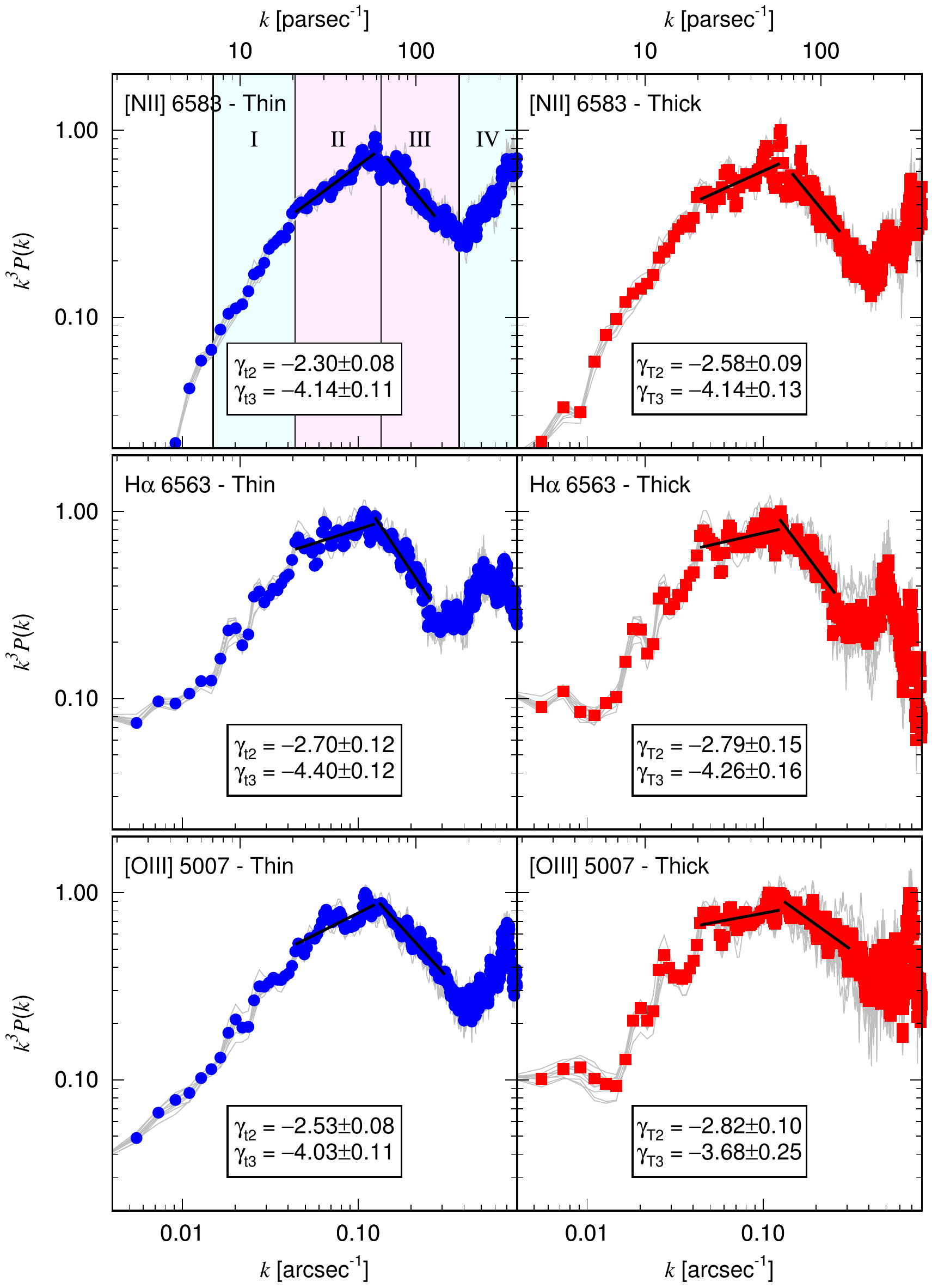}
\caption{Compensated power spectra $k^3P(k)$ of the velocity channels for the
  [\ion{N}{ii}], H$\alpha$ and [\ion{O}{iii}] emission lines for thin (32 velocity
  slices) velocity slices  and thick (one velocity
  slice) in the velocity range $-40$ to $70$~km~s$^{-1}$. Four
  different slope regimes and their approximate wavenumber ranges can
  be identified for each emission line and are indicated approximately
  in the first
  panel only as regimes I, II, III and IV.  The blue
  symbols and red symbols show the combined compensated power spectra for the
  set of 96 different observed longslit data. The grey lines show the
  combined compensated power spectra for a random selection (with replacement) of 96
  observed longslit data. Also shown are the least-squares fits to
  the data points for wavenumber ranges corresponding to power-law indices
  steeper than $-3$ and power-law indices shallower than $-3$ in
  regimes II and III. The errors on the power-law indices come from the variation
of slope among the set of grey lines.}
\label{fig:HNOVCA}
\end{figure*}
\begin{figure*}
\centering
\includegraphics[width=0.75\textwidth]{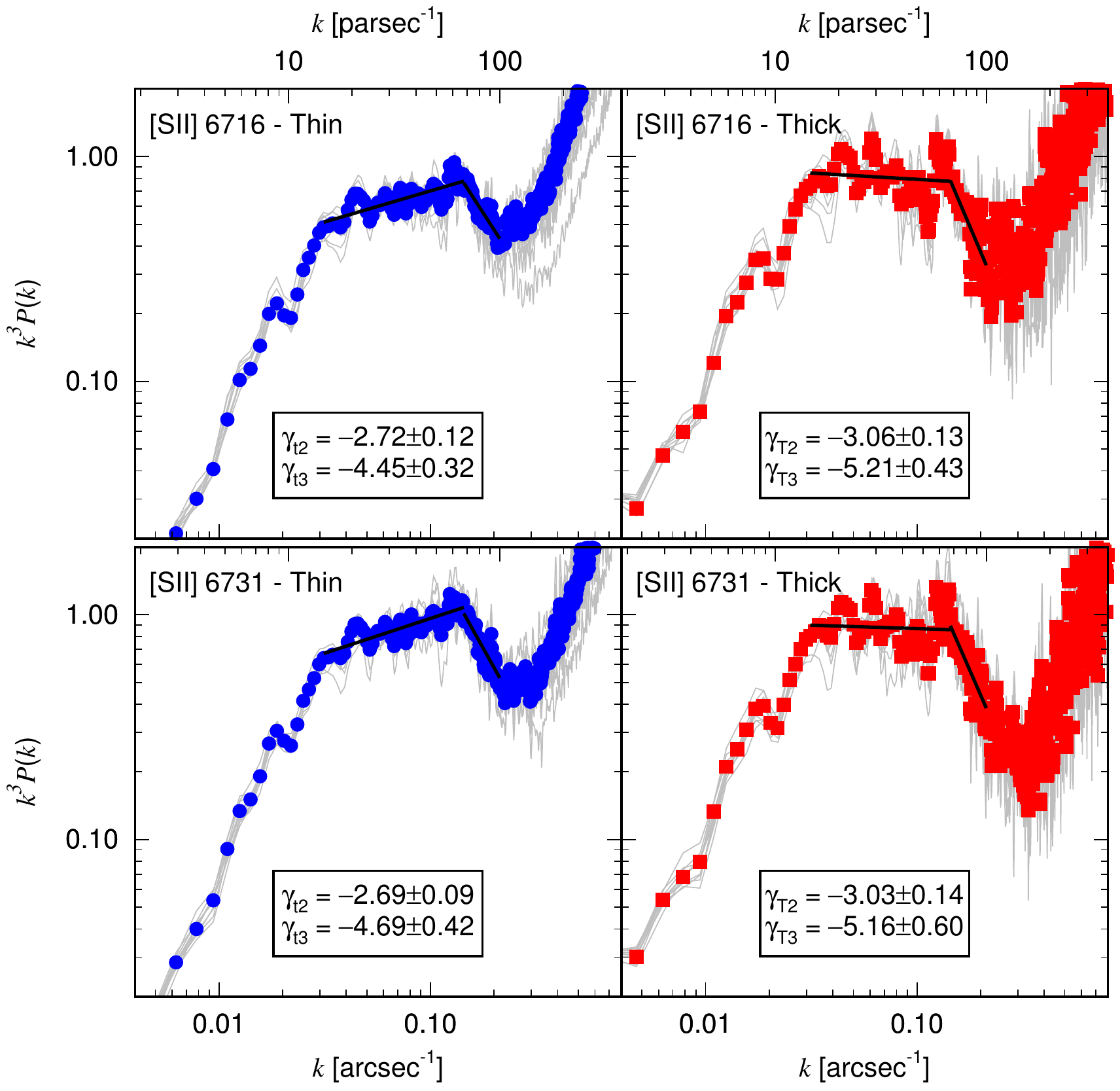}
\caption{Same as Fig.~\protect\ref{fig:HNOVCA} but for the [\ion{S}{ii}] 6716 and
  [\ion{S}{ii}] 6731 emission lines.}
\label{fig:SIIVCA}
\end{figure*}

\begin{figure*}
\centering
\includegraphics[width=0.75\textwidth]{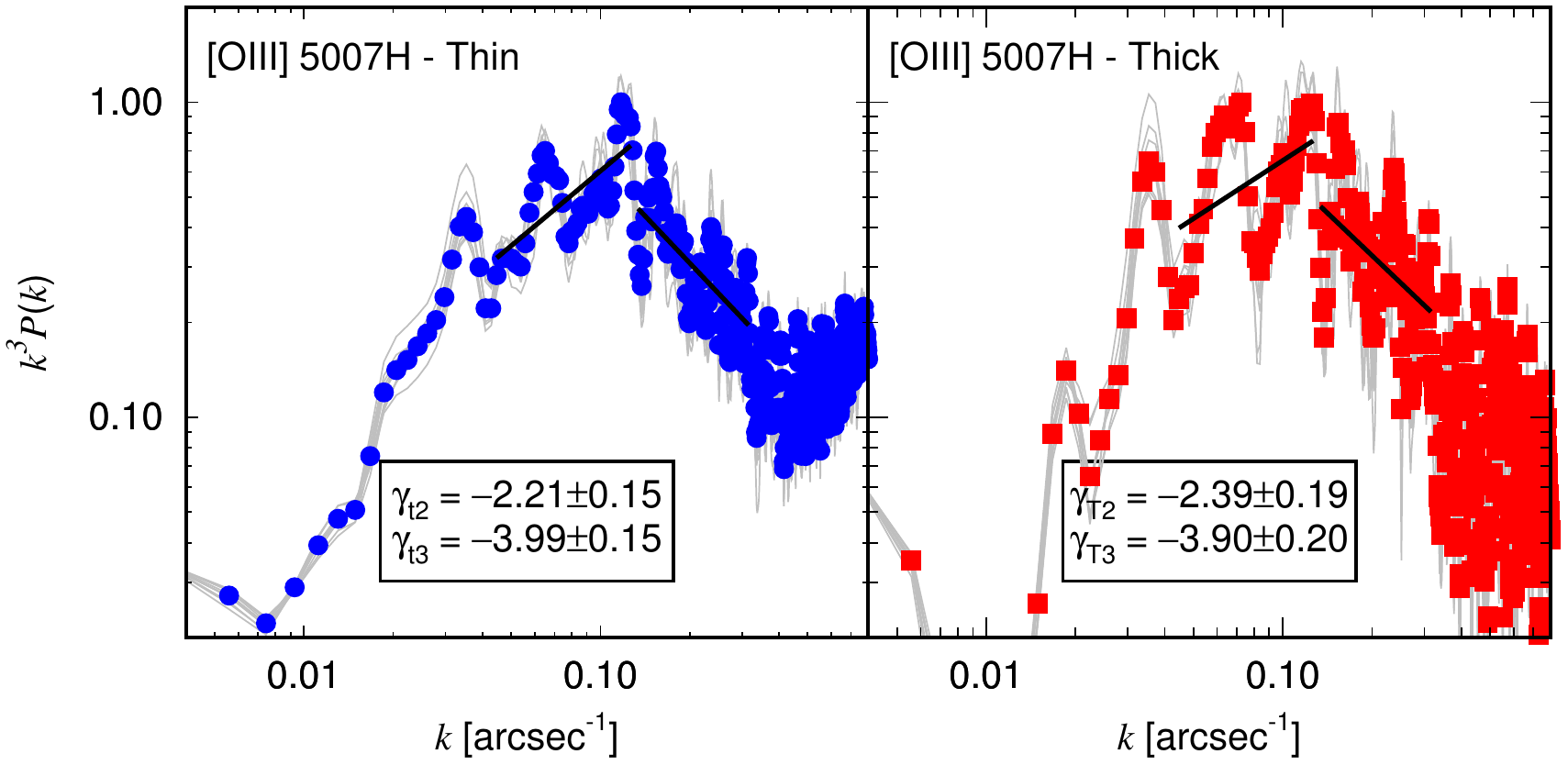}
\caption{Same as Fig.~\protect\ref{fig:HNOVCA} but for the [\ion{O}{iii}] 5007 horizontal slits.}
\label{fig:vcahoriz}
\end{figure*}

The 1D, normalised, compensated power spectra $k^3 P(k)$ for thin
($4$~km\,s$^{-1}$) and thick velocity channels are shown for all
emission lines in Figures~\ref{fig:HNOVCA} and \ref{fig:SIIVCA} and
for the [\ion{O}{iii}] horizontal slits in Figure~\ref{fig:vcahoriz}. The
coloured points represent the average power spectrum of 96 distinct
slits, in the case of the [\ion{N}{ii}]\,$\lambda$6583, H$\alpha$, and
[\ion{O}{iii}]\,$\lambda$5007 emission lines, where the slits cover uniformly
the region shown in the velocity centroid maps. In the
case of the two [\ion{S}{ii}] emission lines, only the 20 highest resolution
slits are used, since the other slits are too affected by noise at
high wavenumber (see Fig.~\ref{fig:siislits} for
the positions of the 20 slits and \citealp{{2008RMxAA..44..181G}} for
observational details). The 18 [\ion{O}{iii}] horizontal slit positions are
shown in Figure~\ref{fig:ohoriz} and the small number of observations
in this dataset means that the compensated power spectra are much
noisier than those of the other datasets.

By plotting $k^3 P(k)$ it
becomes very apparent that there is a break in the power-law behaviour
around wavenumber $k = 0.124$~arcsec$^{-1}$ ($k = 0.144$~arcsec$^{-1}$
for the [\ion{S}{ii}] lines). This corresponds to a size scale of $l \sim 7$ --
$8$~arcsec. Indeed, the behaviour of the power spectrum for all the
emission lines can be divided
into four wavenumber ranges, which we have indicated by I, II, III and
IV in the first panel of Figure~\ref{fig:HNOVCA}. Regimes I and II
correspond to wavenumbers smaller than the break point, while regimes
III and IV correspond to wavenumbers larger than the break point.

For wavenumbers in regime II, the power-law indices of all the power spectra
for all the emission lines are all $\gamma > -3$ for both the thin velocity
channels and the thick channels, with the exception of the
[\ion{S}{ii}] emission lines, where the power law index is $\sim -3$. In all
cases the power laws for the thin channels are less steep than those
of the thick channels. For wavenumbers in regime III, the power-law indices are all steeper than
the critical value 
$\gamma = -3$, indicating that there is very little power at small spatial
scales. Noise dominates the spectra in regime IV. The spectral indices
in regimes I and IV are similar and are both shallow indicating that
the velocity fluctuations at the largest and smallest scales are uncorrelated.

The overall shape of the compensated power spectra are qualitatively
similar for all lines and for both thin and thick slices. In
particular, the break in the slope always occurs at the same scale: 7--8~arcsec, indicating that it must be a feature of the emissivity
fluctuations in the nebula. The power spectrum of the thin slices is
generally shallower (less negative) than that of the thick slices,
which is indicative of the additional effect of velocity fluctuations
(discussed in more depth in \S~\ref{sec:analysis}).

The power spectra resulting from 10 different slit resamplings
are shown in grey in Figures~\ref{fig:HNOVCA} and
\ref{fig:SIIVCA}. The spread in power-law indices we obtain from these
resamplings gives an estimate of the confidence bounds on the spectral
index of the average power spectrum. This spread is not primarily due
to observational uncertainties, which are negligible for all but the
very smallest size scales (largest wavenumbers) but rather to
large-scale inhomogeneity of the power spectrum depending on the
position of the observations.

\subsection{Second-order Structure Functions}
\label{subsec:sf}
\begin{figure*}
\centering
\includegraphics[width=0.9\textwidth]{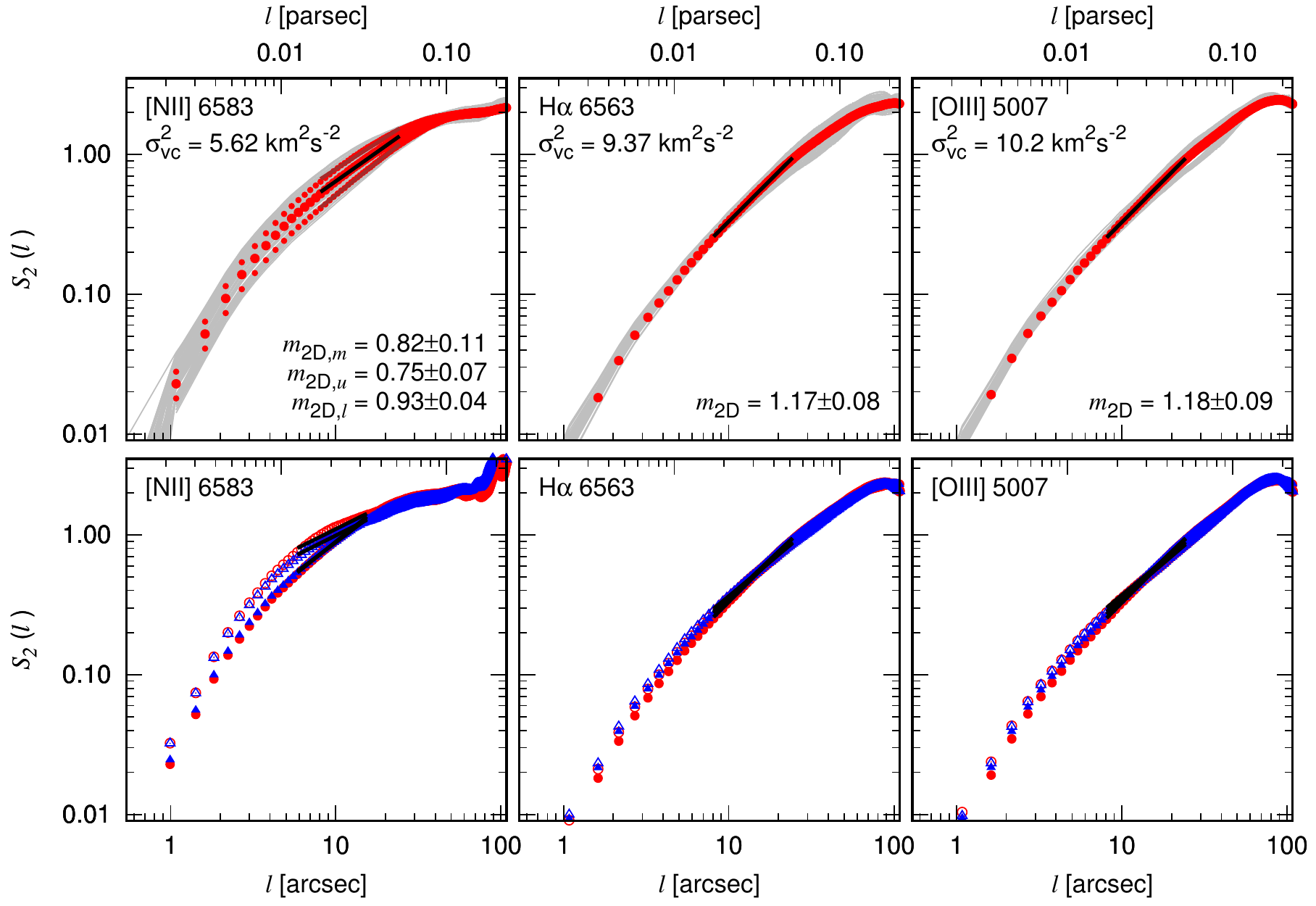}
\caption{Second-order structure functions for the velocity centroid images
  of the [\ion{N}{ii}]\,$\lambda$6583, H$\alpha$ and [\ion{O}{iii}]\,$\lambda$5007
  emission lines considered in this work. Top panel (pdf-selected pixels): the
  structure functions of a random sample of 100 rectangles are shown
  with grey lines, the
  combined structure function is shown  with red circles and the
  power-law fit to the combined structure function is shown with a thick black
  line. The power-law index of the fit and the standard deviation obtained by
  considering fits to the individual rectangle structure functions is
  indicated in each panel. The [\ion{N}{ii}] emission line shows a clear
  bimodal distribution and so fits to the upper and lower parts (60
  and 40 rectangles, respectively) were also obtained separately
  (small red circles). Lower panel: number-of-points weighted
  structure functions (solid symbols) and intensity weighted structure
  functions (open symbols). Red circles
  are for the pdf-selected pixels from the velocity centroid maps
  whereas blue triangles are for the full velocity centroid map with
  no pixel selection. A pixel width
  is equivalent to 0.54$^{\prime\prime}$ in the corresponding velocity
centroid maps from which these structure functions were obtained.}
\label{fig:HNOSF}
\end{figure*}
\begin{figure*}
\centering
\includegraphics[width=0.6\textwidth]{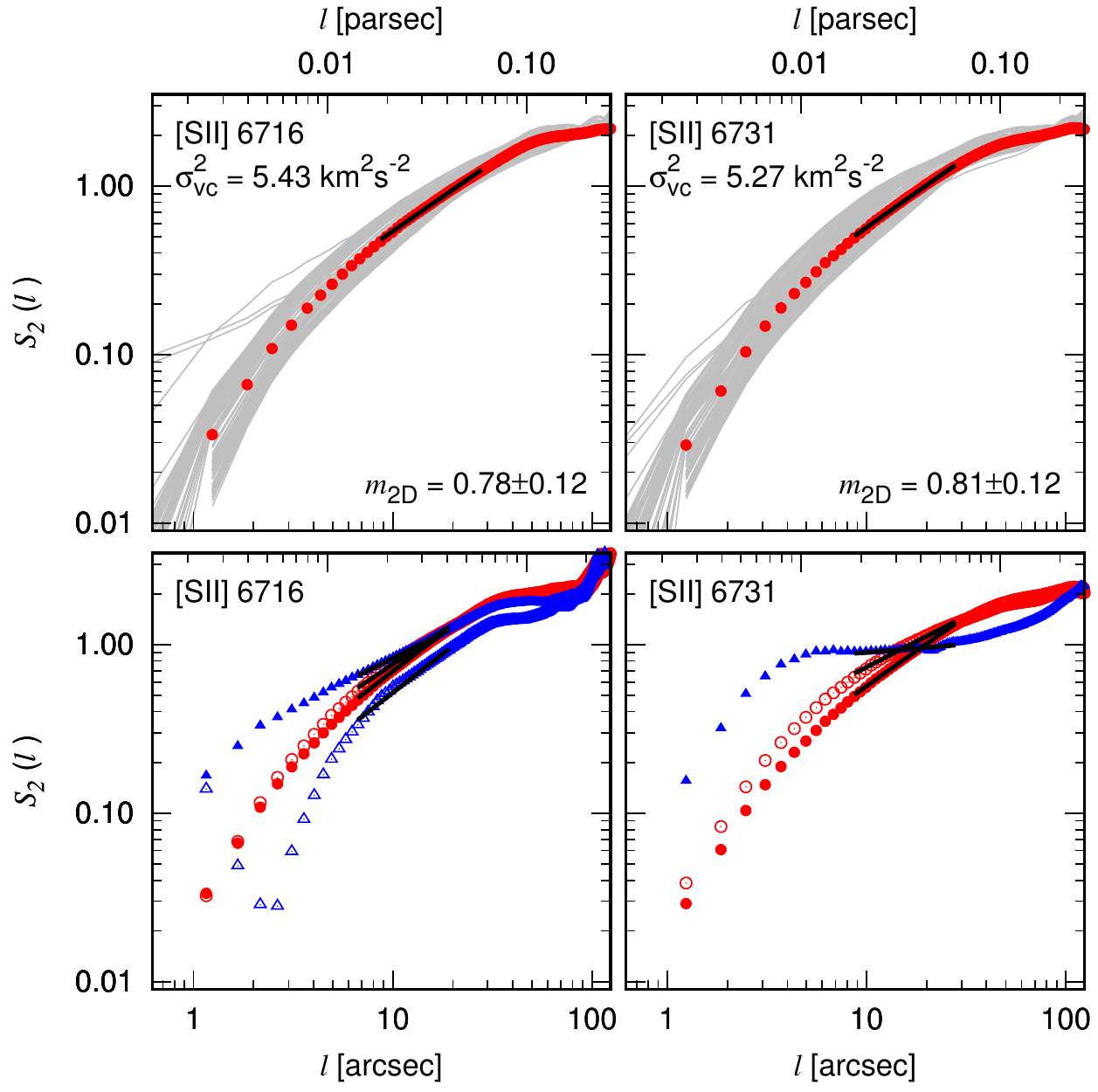}
\caption{Same as Fig.~\protect\ref{fig:HNOSF} but for the [\ion{S}{ii}]\,$\lambda$6716
  and [\ion{S}{ii}]\,$\lambda$6731 emission lines. A pixel width is
  equivalent to 0.62$^{\prime\prime}$ in the corresponding velocity
  centroid maps from which these structure functions were obtained. }
\label{fig:SIISF}
\end{figure*}

\subsubsection{Shape and Slope of Structure Function for each emission
  line}

We use Equation~\ref{eq:sf} to calculate the second-order structure
functions of the velocity centroid maps for the [\ion{N}{ii}]\,$\lambda$6583, H$\alpha$,
[\ion{O}{iii}]\,$\lambda$5007, [\ion{S}{ii}]\,$\lambda$6716 and
[\ion{S}{ii}]\,$\lambda$6731 emission lines.  Figures~\ref{fig:HNOSF} and
\ref{fig:SIISF}  (top panels) show the complete structure function of
the pdf-selected pixels, together with
the structure functions obtained from 100 smaller rectangular frames
within the velocity centroid map (see Fig.~\ref{fig:frames}). The fit to the power-law index of
the structure function is performed over the spatial separations
corresponding to regime II of the velocity channel analysis (roughly $8 < l < 22$~arcsec for the [\ion{N}{ii}],
H$\alpha$ and [\ion{O}{iii}] emission lines, and $7 < l < 32$~arcsec for [\ion{S}{ii}]
emission lines; see Fig~\ref{fig:HNOVCA}). The noise, estimated to be
equal to the one-pixel separation value $S_2(1)$, has been
subtracted from the structure function. In the figures, we only plot
the structure function up to spatial separations $l = 120$~arcsec. Above this
value, the structure function falls away steeply.

Unlike in the case of the power spectra, the structure functions show
no clear break at a scale of $\sim 8$~arcsec. The power-law fit to
scales $> 8$~arcsec is not too bad a fit to smaller scales
also. The structure function has a slight negative curvature, giving gradually
steeper slopes at smaller scales. The more pronounced steepening below
$2$~arcsec is due to the spatial resolution of the observations and
corresponds to the region where the seeing and
inter-slit separation become important.

The steepest power-law indices over the fit range are for the
H$\alpha$ and [\ion{O}{iii}] emission lines. The form of the structure
function curves is the same for both emission lines and the value of
the power-law index is $m_\mathrm{2D} \sim 1.2 \pm 0.1$ in both
cases. The two [\ion{S}{ii}] emission lines have very similar structure
functions to each other. Their slopes are shallower than those of the H$\alpha$
and [\ion{O}{iii}] structure functions, having $m_\mathrm{2D} \sim 0.8 \pm
0.1$. There is considerable variation at the smallest scales in the structure function among
the different randomly selected frames. Finally, the structure functions for the
[\ion{N}{ii}] line can be divided into two distinct populations: one set
corresponds to frames selected from the upper part of the velocity
centroid map, the other to frames from the lower half (see
Figure~\ref{fig:frames}). One set has a power-law index similar to that
of the [\ion{S}{ii}] lines, while  the other is intermediate between the [\ion{S}{ii}]
and H$\alpha$ cases.

In the lower panels of Figures~\ref{fig:HNOSF} and \ref{fig:SIISF} we
show alternative structure functions for the same emission-line
velocity centroid maps. The solid red circles show the same
pdf-selected structure functions as in the upper panels, while the
solid blue triangles show the structure functions obtained when all
the pixels are used. There is little difference between the two
structure functions for the [\ion{N}{ii}], H$\alpha$ and [\ion{O}{iii}] emission lines
but the [\ion{S}{ii}] structure functions show considerable differences. This
is because more pixels are eliminated from the velocity centroid maps on the basis of the pdf analysis
for the [\ion{S}{ii}] lines: some of these pixels correspond to high-velocity
features, such as HH objects, while others correspond to numerical
noise. Evidently, these pixels influence the structure function. 

Also
in the lower panels of Figures~\ref{fig:HNOSF} and \ref{fig:SIISF}, we
plot the intensity weighted structure functions of the corresponding
cases. 
The figures show
that
once again, for the H$\alpha$ and [\ion{O}{iii}] emission lines, there is
little difference between these new structure functions (open symbols) and the
number-of-points-weighted structure functions (solid symbols). However, this time the
[\ion{N}{ii}] structure functions are different, with the intensity weighting
producing structure functions with shallower slopes (smaller power-law
indices). For the shorter
wavelength [\ion{S}{ii}]\,$\lambda$6716 case, the intensity weighting makes
little difference to the pdf-selected structure function (red circles)
but does
reduce the noise contribution to the full structure function (blue
triangles) and leads
to a similar power-law index to the pdf-selected cases over the fitted
range. For the longer wavelength [\ion{S}{ii}]\,$\lambda$6731 case, the
intensity weighting produces a shallower power law, even for the
pdf-selected case (red circles). It is not possible to plot the
intensity weighted structure function for the full set of pixels on
the same scale because the contributions of regions of bad pixels both in the
velocity centroid map and the intensity map dominate
 at all scales.

\subsection{Analysis of power-law indices}
\label{sec:analysis}
\begin{table*}
\caption{Description and Relationships between Power-Law Indices}
\begin{tabular}{lcccc}
\hline
Description & Power-law & Relationship && Kolmogorov \\
                   & Index         &                     && Value \\
\hline
3D emissivity fluctuations & $n_\mathrm{E}$ && &  \\
3D velocity fluctuations & $n$ & $n = -3 - m_\mathrm{3D}$ &\textit{(a)}&$-11/3$ \\
3D second-order structure function & $m_\mathrm{3D}$ & $m_\mathrm{3D} = -3 - n$&\textit{(a)}& $2/3$ \\
2D second-order structure function & $m_\mathrm{2D}$ &&&\\
\hspace*{1cm}\textit{Projection Smoothing}& & $m_\mathrm{2D} = m_\mathrm{3D}+1$&\textit{(b)}&$5/3$ \\ 
\hspace*{1cm}\textit{Sheetlike Emission}& & $m_\mathrm{2D} \sim m_\mathrm{3D}$&\textit{(c)}&$2/3$ \\ 
Intensity fluctuations & $\gamma$ & && \\
\hspace*{.5cm}Very Thick Velocity Slice &
$\gamma_\mathrm{T}$&  $\gamma_\mathrm{T} \sim n_\mathrm{E}$ &\textit{(d)}& \\ 
\hspace*{.5cm}Thin Velocity Slice &$\gamma_\mathrm{t}$ &&& \\
\hspace*{1cm}\textit{Shallow density} $n_\mathrm{E} > -3$ & & $\gamma_\mathrm{t} = \gamma_\mathrm{T} +
m_\mathrm{3D}/2$ &\textit{(d)}& \\ 
\hspace*{1cm}\textit{Steep density} $n_\mathrm{E} < -3$ & &
$\gamma_\mathrm{t}  = -3 + m_\mathrm{3D}/2$ &\textit{(d)}&  $-8/3$ \\
\hline
\multicolumn{5}{l}{\textit{(a)} \citet{1941DoSSR..30..301K} }\\
\multicolumn{5}{l}{\textit{(b)} \citet{1951ZA.....30...17V}  }\\
\multicolumn{5}{l}{\textit{(c)} \citet{1987ApJ...315L..55C}}\\
\multicolumn{5}{l}{\textit{(d)} \citet{2000ApJ...537..720L}}\\
\end{tabular}
\label{tab:summary}
\end{table*}
\begin{table*}
\caption{Regime II spectral indices and relationships}
\begin{tabular}{lcccccc}
\hline
Emission & Thin & Thick & Steep or
&\multicolumn{2}{c}{$m_\mathrm{3D}$} & $n$  \\
Line &$\gamma_\mathrm{t2}$& $\gamma_\mathrm{T2}$ & Shallow &
$2(\gamma_\mathrm{t2} -
\gamma_\mathrm{T2})$&$2(\gamma_\mathrm{t2} + 3)$ & $-3-m_\mathrm{3D}$\\
\hline
$\left[\mbox{SII}\right] \lambda$6716 &  $-2.72\pm 0.12$&$-3.06\pm
0.13$  &Steep  & \dots  &$0.56\pm 0.24$  & $-3.56\pm 0.24$  \\
$\left[\mbox{SII}\right] \lambda$6731 &  $-2.69\pm 0.09$&$-3.03\pm
0.14$  &Steep  & \dots  &$0.62\pm 0.22$  & $-3.62\pm 0.22$   \\
$\left[\mbox{NII}\right] \lambda$6583&  $-2.30\pm 0.08$&$-2.58\pm
0.09$  &Shallow  & $0.56\pm 0.24$  & \dots  & $-3.56\pm 0.24$  \\
H$\alpha$\,$\lambda$6563 &  $-2.70\pm 0.12$&$-2.79\pm 0.15$ &Shallow  &\dots &\dots  & \dots \\
$\left[\mbox{OIII}\right] \lambda$5007 &  $-2.53\pm 0.08$&$-2.82\pm
0.10$  &Shallow  & $0.58\pm 0.26$  &\dots  & $-3.58\pm 0.26$    \\
$\left[\mbox{OIII}\right] \lambda$5007H &  $-2.21\pm 0.15$&$-2.39\pm
0.19$  &Shallow  &$0.36\pm 0.24$  & \dots & $-3.36\pm 0.24$  \\
\hline
\end{tabular}
\label{tab:vcafits2}
\end{table*}
\begin{table*}
\caption{Regime III spectral indices and relationships}
\begin{tabular}{lccccc}
\hline
Emission & Thin & Thick & Steep or &$m_\mathrm{3D}$ & $n$ \\
Line &$\gamma_\mathrm{t3}$& $\gamma_\mathrm{T3}$ & Shallow &
$2(\gamma_\mathrm{t3} + 3)$ & $-3-m_\mathrm{3D}$\\
\hline
$\left[\mbox{SII}\right] \lambda$6716 &$-4.45\pm 0.32$  &$-5.21\pm
0.43$  &Steep    & NEG  &  N/A \\
$\left[\mbox{SII}\right] \lambda$6731 &$-4.69\pm 0.42$  &$-5.16\pm 0.60$  &Steep  & NEG &  N/A  \\
$\left[\mbox{NII}\right] \lambda$6583&$-4.14\pm 0.11$  &$-4.14\pm
0.13$  &Steep   &  NEG &  N/A \\
H$\alpha$\,$\lambda$6563 &$-4.40\pm 0.12$  &$-4.26\pm 0.16$ &Steep
&  NEG & N/A \\
$\left[\mbox{OIII}\right] \lambda$5007 &$-4.03\pm 0.11$  &$-3.68\pm
0.25$  &Steep   &  NEG & N/A   \\
$\left[\mbox{OIII}\right] \lambda$5007H &$-3.99\pm 0.15$  &$-3.90\pm
0.20$  &Steep   & NEG & N/A \\
\hline
\end{tabular}
\label{tab:vcafits3}
\end{table*}
\begin{table}
\caption{Structure function power-law indices and relationships}
\begin{tabular}{lccc}
\hline
Emission & $m_\mathrm{2D}$ &\multicolumn{2}{c}{$m_\mathrm{3D}^*$}  \\
Line& & $m_\mathrm{2D}-1$ & $m_\mathrm{2D}$ \\
\hline
$\left[\mbox{SII}\right] \lambda$6716 & $0.78\pm 0.12$ &\dots &$0.78\pm 0.12$\\
$\left[\mbox{SII}\right] \lambda$6731 & $0.81\pm 0.12$ &\dots
&$0.81\pm 0.12$ \\
$\left[\mbox{NII}\right] \lambda$6583 & $0.82\pm 0.11$ &\dots & $0.82\pm 0.11$\\
H$\alpha$\,$\lambda$6563 & $1.17\pm 0.08$& $0.17\pm 0.08$ & \dots \\
$\left[\mbox{OIII}\right] \lambda$5007 &$1.18\pm 0.09$ & $0.18\pm
0.09$&\dots \\
\hline
\end{tabular}
\label{tab:sfind}
\end{table}

We want to use the power-law indices from the velocity channel
analysis and the second-order structure functions obtained from the
spectroscopic observations to recover the three-dimensional velocity
statistics of the ionized gas in the Orion Nebula. There are various
relationships between the observationally derived power-law indices
and the underlying 3D density and velocity fluctuations, which are
summarized in Table~\ref{tab:summary} and described below.

The velocity channel analysis gives us the power-law indices of the
average power spectra of thin and thick velocity slices. The thickest
velocity slice corresponds to the velocity-integrated surface
brightness and the power-law indices of the spectra of the thick
slices are predicted to be equal to the those of the 3D power
spectra of the respective emissivities, i.e., $\gamma_\mathrm{T} =
n_\mathrm{E}$ \citep{2000ApJ...537..720L}. Thus, we can use our calculated
value of $\gamma_{\mathrm{T}}$ as an estimate of
$n_\mathrm{E}$.  The
critical value $n_\mathrm{E} \sim -3$ divides `steep' from `shallow'
power spectra.  A `steep' emissivity spectrum ($n_\mathrm{E} < -3$)
means that the emissivity is dominated by fluctuations at intermediate to
large scales, while a `shallow' spectrum ($n_\mathrm{E} > -3$)
indicates that the emissivity is dominated by fluctuations at small
scales. As the velocity slices become thinner, velocity fluctuations
dominate the spectra. By using a combination of thick slice and thin
slice spectral indices, we can, in theory recover the velocity
power-law index.

There are relations between the spectral index of the 3D velocity
fluctuation power spectrum and the power-law indices of the VCA power
spectra, depending on whether the power spectrum of the emissivity
fluctuations is `steep' or `shallow'
\citep{2000ApJ...537..720L}. 
In the steep case, the power-law index of the average power spectrum of
the thin isovelocity channels is given by
\begin{equation}
\gamma_\mathrm{t} = -3 + \frac{1}{2} m_\mathrm{3D} \ , 
\label{eq:steeprel}
\end{equation}
where $n = -3 - m_\mathrm{3D}$ is the spectral index of the 3D
velocity power spectrum (see Table~\ref{tab:summary}). For the `shallow' case, the relation is with the difference
in the power-law indices of the average power-spectra of the thin and
thick velocity slices:
\begin{equation}
 \gamma_\mathrm{t} - \gamma_\mathrm{T} = \frac{1}{2} m_\mathrm{3D} \ .
\label{eq:shallowrel}
\end{equation}

There are also
relationships between the power-law indices of the two-dimensional
structure function of the emission-line velocity centroid map and the
three-dimensional structure function of the velocity field. For
projection from 3 to 2 dimensions we have $m_\mathrm{2D} =
m_\mathrm{3D} + 1$, which is known as projection smoothing \citep{{1951ZA.....30...17V},{1958RvMP...30.1035M},{1987ApJ...317..686O},{2004ApJ...604..196B}}. If,
however, the distribution of emitters is sheet-like, i.e., essentially
two-dimensional, then $m_\mathrm{2D} \approx m_\mathrm{3D}$
\citep{{1987ApJ...315L..55C},{1995ApJ...454..316M}}. The three-dimensional structure function
power-law index is related to the three-dimensional power
spectrum spectral index ($n$) of the underlying  velocity field through
$m_\mathrm{3D} = -3 - n$, where $n = -\frac{11}{3}$ for homogeneous,
incompressible turbulence (i.e., Kolmogorov spectrum) (see Table~\ref{tab:summary}).

\subsubsection{VCA: Regime II}
\label{sssec:VCAR2}
For the low
wavenumber range, $k < 0.124$~arcsec$^{-1}$, corresponding to spatial scales $> 8$~arcsec, the
power-law indices of the thick slice power spectra for regime II are
$\gamma_{\mathrm{T2}}$ and are listed in Table~\ref{tab:vcafits2}.
We consider each emission line in turn, beginning with the [\ion{S}{ii}] lines,
which represent the lowest ionization state of our data. $\gamma_\mathrm{T} = -3.06$ and
$\gamma_\mathrm{T} = -3.03$ for the [\ion{S}{ii}]\,$\lambda$6716 and
[\ion{S}{ii}]\,$\lambda$6731 lines, respectively. This is the critical value
separating `steep' from  `shallow' power spectra. Using the relations
between power-law indices listed in Table~\ref{tab:summary} (see also Eq.~\ref{eq:steeprel}), we obtain
$m_\mathrm{3D} = 0.56 \pm 0.24$
and $m_\mathrm{3D} = 0.62 \pm 0.22$ for the short and long
wavelength [\ion{S}{ii}] lines, respectively. The corresponding power-law
indices of the 3D velocity fluctuations are then $n = -3.56 \pm 0.24$
and $n = -3.62 \pm 0.22$, respectively. These values of $n$ are consistent with the Kolmogorov value of
$n = -3.67$. Note that the [\ion{S}{ii}] lines are least affected by thermal broadening
and, moreover, the 20 spectra used to calculate the power spectra have
the highest velocity resolution of all our observational data.

Moving to higher ionization lines, the calculated `thick' and `thin'
slice power-law indices of the [\ion{N}{ii}]\,$\lambda$6583 spectra fall into
the `shallow' regime. Following the same analysis as above, we recover
$n = -3.56 \pm 0.24$, which is again close to the Kolmogorov
value. For the H$\alpha$ power spectra, there is little difference
between the thin and thick power-law indices since thermal broadening
is important for this line (see, e.g., \citetalias{2014MNRAS.445.1797M}). It is therefore not possible to recover
a meaningful value of $n$. Finally, the [\ion{O}{iii}]\,$\lambda$5007
power spectra again fall into the `shallow' regime and we obtain $n =
-3.58 \pm 0.26$.  

It is gratifying that the [\ion{N}{ii}], [\ion{O}{iii}] and [\ion{S}{ii}] power spectra
power-law indices are consistent with each other and with the
Kolmogorov value, within the errors.

\subsubsection{VCA: Regime III}
\label{ssec:vcar3}

For the high wavenumber range $k > 0.124$~arcsec$^{-1}$, corresponding
to spatial scales $< 8$~arcsec,
only the [\ion{S}{ii}]\,$\lambda$6716 and [\ion{S}{ii}]\,$\lambda$6731 power spectra
have `thick' slice  power-law indices steeper than their `thin' slice
power-law indices. The [\ion{N}{ii}], H$\alpha$ and [\ion{O}{iii}] have curiously opposite
behaviour and their power-law indices cannot be accommodated by the
theory (see Table~\ref{tab:vcafits3}). For the [\ion{S}{ii}] power spectra, the value of $\gamma_\mathrm{T}$
falls into the `steep' regime. However, for both [\ion{S}{ii}] lines, the value of
$\gamma_\mathrm{t}$ is less than $-3$ and so this would lead to negative values
for $m_\mathrm{3D}$ on applying Equation~\ref{eq:steeprel}. 

\subsubsection{Structure function power indices}
\label{sssec:SFPI}
We fit the structure functions over the separation range corresponding
to VCA regime II. The structure function graphs themselves do not strongly suggest that
any particular range is preferable, although 
in the range $l <  4$~arcsec the structure function falls off very
steeply due to the effects of seeing and the interslit separation,
which has to be interpolated to produce the velocity centroid maps.
At the largest separations, the structure function power-law index flattens off, i.e
$m_\mathrm{2D} \sim 0$ for $l > 100$~arcsec. Over the separation range we fit,
  we find $m_\mathrm{2D} \sim 1.2$ for the H$\alpha$
  and [\ion{O}{iii}] lines, and $m_\mathrm{2D} < 1$ for the [\ion{S}{ii}] and [\ion{N}{ii}] lines.

  The values we obtain for $m_\mathrm{2D}$ allow us to determine a
  range for $m_\mathrm{3D}$. The 3D structure-function index must lie
  between the two limits given by projection smoothing and a
  sheet-like distribution of emitters, that is $m_\mathrm{2D}-1 <
  m_\mathrm{3D} < m_\mathrm{2D}$ (see
  Table~\ref{tab:summary}). Our lowest value is $m_\mathrm{2D} \simeq
  0.80$ and the highest is $m_\mathrm{2D} \simeq 1.18$, therefore
  $m_\mathrm{3D}$ must lie in the range $0.18 < m_\mathrm{3D} < 0.80$. This is consistent with
  the results of the VCA, for which we found $m_\mathrm{3D} \simeq
  0.67$ but is such a wide range of values that it is not a useful diagnostic.

\subsection{Line-of-sight velocity dispersions}
\label{sec:poslos}

\begin{figure}
  \centering
  \includegraphics[width=\linewidth]{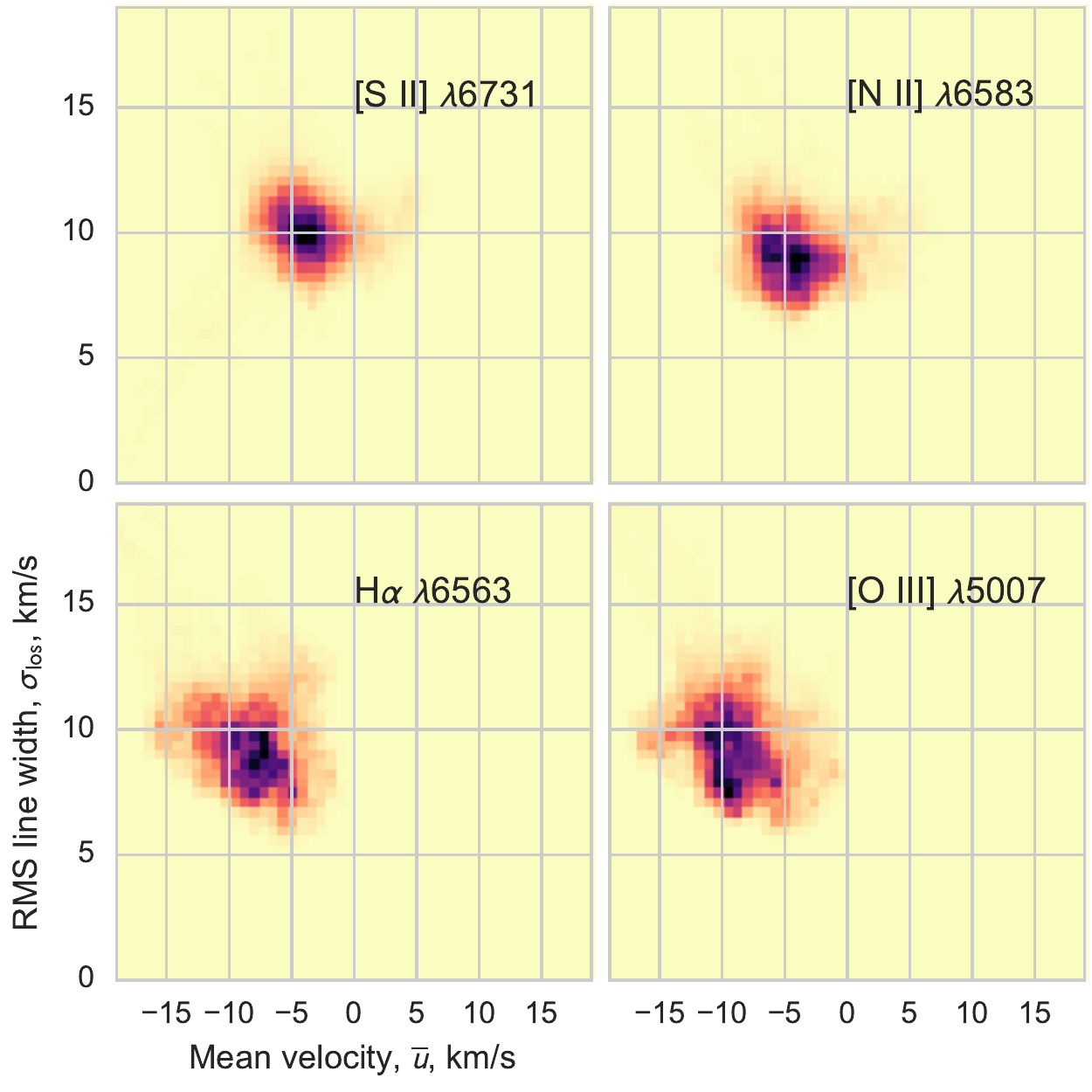}
  \caption{Joint distribution of moment-derived line-of-sight velocity
    dispersion and mean velocity.  The thermal, instrumental, and
    fine-structure contributions to the velocity dispersion have been
    subtracted in quadrature.  Mean velocities are shown with respect
    to the systemic velocity of the stellar cluster (\(+7~\kms\) in
    the Local Standard of Rest frame, or \(+25~\kms\) in the
    Heliocentric frame, \citealp{Tobin:2009a}).}
  \label{fig:observed-vmean-sigma}
\end{figure}
The spectral lines are broadened beyond the thermal component by a
combination of ordered and disordered motions along the line of
sight. Moreover, the lines often do not have Gaussian profiles and can
sometimes be separated into distinct components that could originate
from different physical components of the ionized gas in the nebula
having different kinematic characteristics \citep{1988ApJS...67...93C}. 
Contributions to ordered motions include expansion, contraction and
rotation, while random motions are often ascribed loosely to ``turbulence''.

The non-thermal line-of-sight velocity dispersion 
\(\sigma\los\)
may be estimated by
correcting the observed line widths for the contributions from the
spectrograph resolution, 
fine-structure splitting, 
and thermal Doppler broadening, 
as in Eq.~(2) of \citet{2008RMxAA..44..181G}.  
A striking fact about the nebular line profiles is that
the line-of-sight velocity dispersion is several times larger than the
plane-of-sky dispersion in mean velocities.  This is illustrated in
Figure~\ref{fig:observed-vmean-sigma}, which shows flux-weighted
histograms of the non-thermal RMS line widths versus mean velocity for
all lines of the \citet{2008RMxAA..44..181G} spectral atlas.   The
line-of-sight RMS velocity dispersion is 9--10~\kms, whereas the RMS
plane-of-sky dispersion of the mean velocities is only 2--4~\kms.  

\begin{figure}
  \centering
  \includegraphics[width=\linewidth]{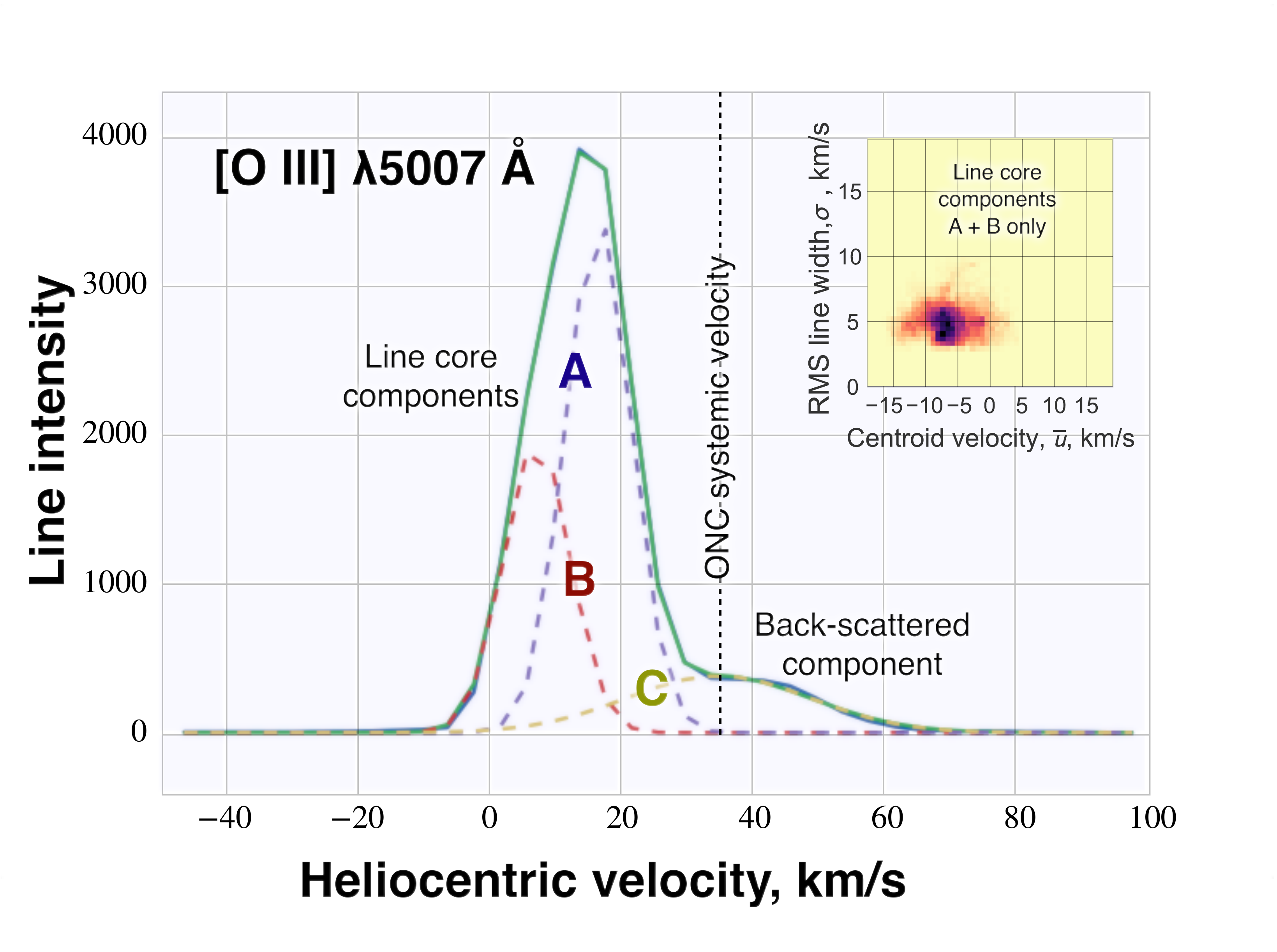}
  \caption{Typical [\ion{O}{iii}] line proile, showing Gaussian
    decomposition into three components, following the labeling of
    \citet {1988ApJS...67...93C}.  The red-shifted component~C is due to
    back-scattering of the blue-shifted components A and B by dust
    that lies behind the emitting gas.  Inset figure is same as lower
    right panel of 
    Fig.~\ref{fig:observed-vmean-sigma}, but after removing
    Component~C from the profile.}
  \label{fig:gauss}
\end{figure}

However, the observed line widths are affected by additional
broadening due to dust-scattering \citep{{1988ApJS...67...93C},
  {Henney:1998a}}, which gives an extended red
wing to the line profile that contains 10--20\% of the total line
flux (see Fig.~\ref{fig:gauss}).   By means of fitting multiple
Gaussian components to each line profile, it is possible to
effectively remove this scattered component and calculate statistics
for the line core alone.  We have done this for the \oiii{} line, with
results shown in the inset box of Figure~\ref{fig:gauss}.  It can be
seen that the line-of-sight velocity dispersion is reduced by almost a
factor of two.  

\begin{figure}
  \centering
  \includegraphics[width=\linewidth]{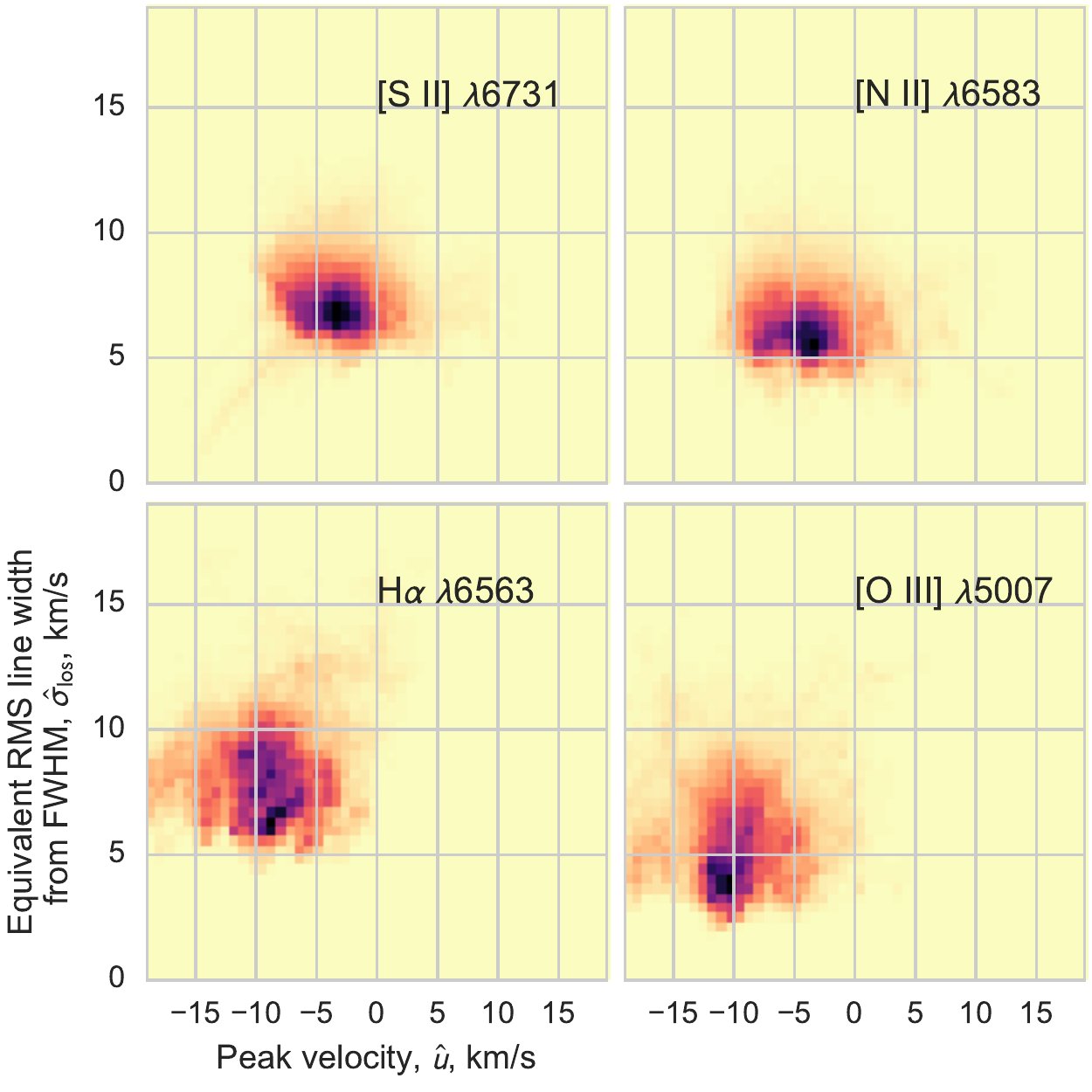}
  \caption[]{Same as Fig.~\ref{fig:observed-vmean-sigma} but
    substituting the FWHM-derived line width \(\hat{\sigma}\los\) for the
    moment-derived line width \(\sigma\los\) and substituting the peak
    velocity \(\hat{u}\) for the mean velocity \(\bar{u}\).  }
  \label{fig:observed-vmean-sigma-fwhm}
\end{figure}

It is more difficult to perform the Gaussian decomposition for the
lower ionization lines since the back-scattered component is not as
cleanly separated from the core component.  On the other hand, an
alternative method of suppressing the effects of scattering on the
line widths is to use the Full Width at Half Maximum (FWHM) instead of
the RMS width.  Even a weak component with a velocity that is very
different from the line centroid can have a large effect on the
\(\sigma\los\) due to the \((u - \bar{u})^2\) dependence of the second
velocity moment.  But the same component will have almost no effect on
the FWHM if its amplitude is less than half that of the line core.  We
therefore define a FWHM-based effective \(\sigma\) as
\(\hat{\sigma}\los = \mathrm{FWHM} / \sqrt{8 \ln 2} \approx
\mathrm{FWHM} / 2.355\), with the constant of proportionality being
chosen so that \(\hat{\sigma}\los = \sigma\los\) for a Gaussian profile.  The
results are shown in Figure~\ref{fig:observed-vmean-sigma-fwhm}, where
it can be seen that nearly all the lines show substantial reductions
in the line-of-sight line widths with respect to the moment-derived
values of Figure~\ref{fig:observed-vmean-sigma}.  The exception is
\ha{}, where larger thermal broadening means that \(\hat{\sigma}\los\) is
still contaminated by the back-scattered component due to blending.
Note that \(\hat{\sigma}\los\) is insensitive to not only the scattered
component, but also to any other weak component that is not blended
with the line core.  For the low-ionization lines such as \nii{} and
\sii{}, this includes the kinematic component known as the Diffuse
Blue Layer \citep{Deharveng:1973a}, which produces line splitting in
the SE and N regions of our observed maps.  

\begin{figure*}
  \centering
  \begin{tabular}{@{}l@{}l@{}}
    (\textit{a}) & (\textit{b}) \\
    \includegraphics[width=0.48\linewidth]{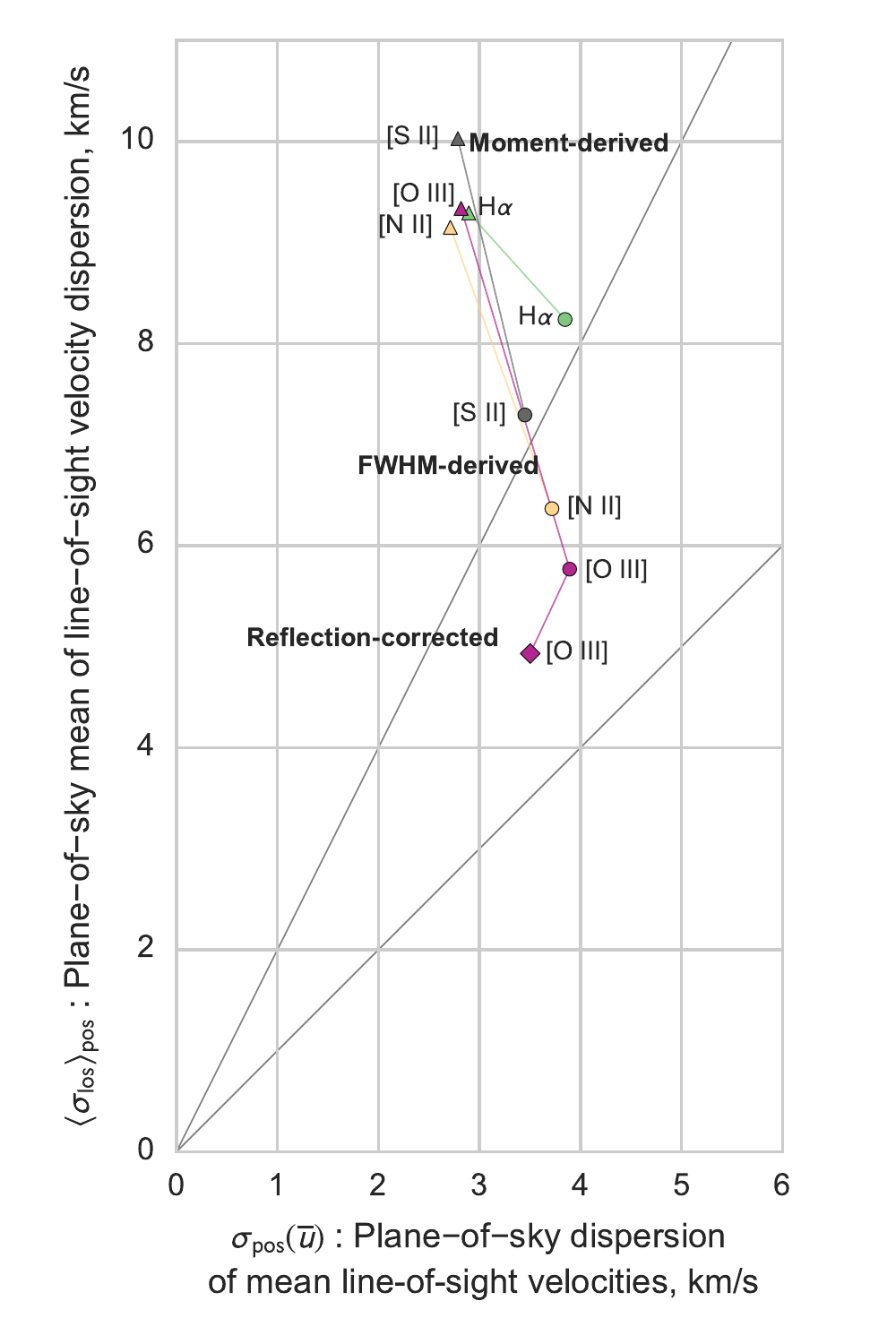} 
    & \includegraphics[width=0.48\linewidth]{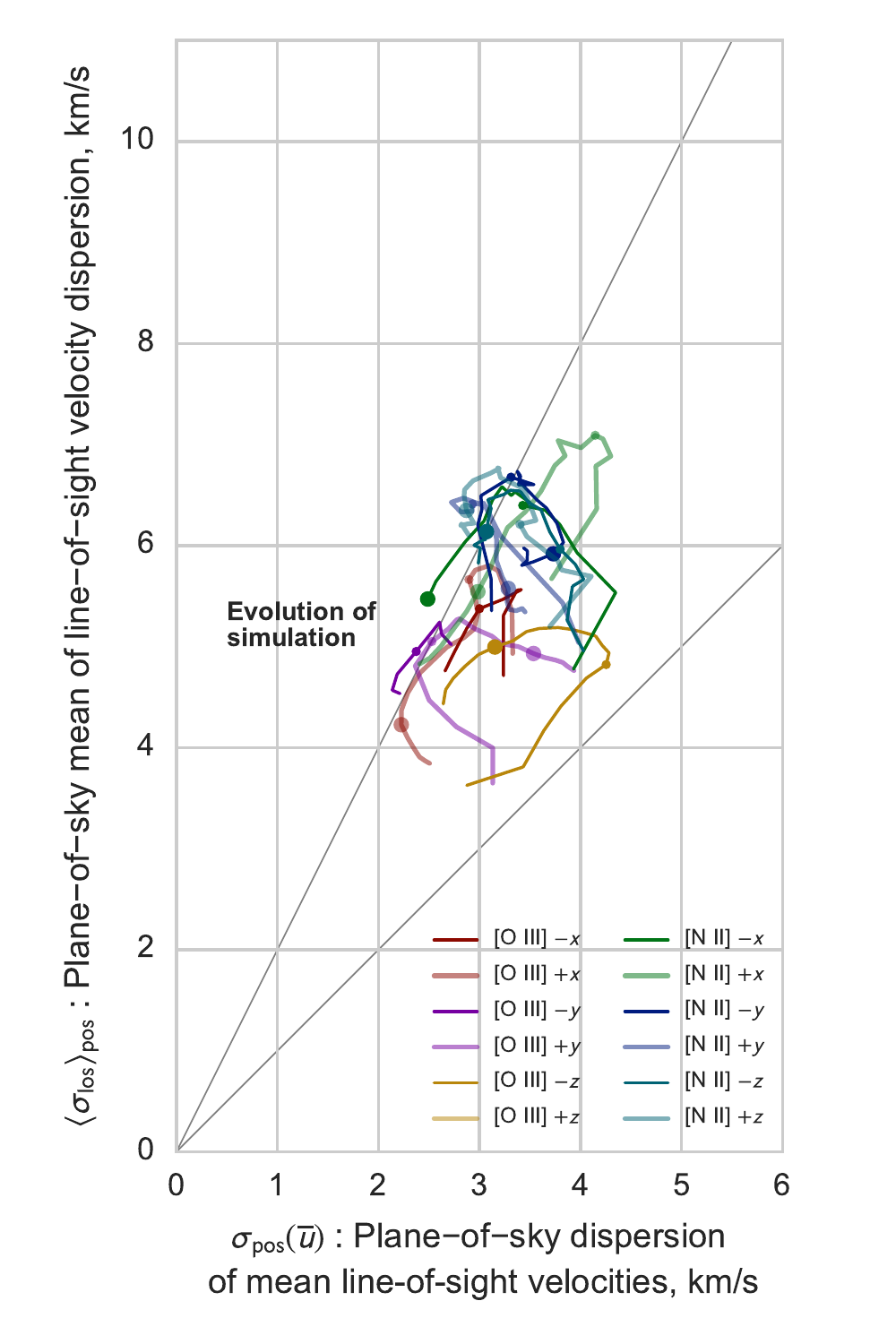} \\
  \end{tabular}
  \caption{(\textit{a})~Observational comparison of mean line-of-sight
    velocity dispersion (vertical axis) with plane-of-sky dispersion
    of mean line-of-sight velocity (horizontal axis).  All averages
    were performed with flux weighting.  Triangle symbols show the
    moment-derived values, which are contaminated by the
    back-scattered component.  Circle symbols show the more robust
    FWHM-derived values, while the diamond symbol shows the
    reflection-corrected result from multi-Gaussian fitting, which was
    only possible for \oiii.  The diagonal grey lines show the
    relationships \(\sigma\los = \sigma\pos\) and \(\sigma\los =
    2\sigma\pos\).  (\textit{b})~The same quantities derived
    from evolutionary tracks of the turbulent \hii{} region simulation
    described in \citetalias{2014MNRAS.445.1797M}  for different viewing directions
    as shown in the key: blue/green lines are for \nii{}, while
    red/purple/yellow lines are for \oiii{}. Evolutionary times of
    0.15~Myr (small dot) and 0.25~Myr (larger dot) are marked on each
    track, and only those times where the mean line velocity is
    blueshifted are shown (see Fig.~\ref{fig:simulation-stats-evo} for
    the data used here).  }
  \label{fig:obs-sigma-sigma}
\end{figure*}

\section[]{Discussion}
\label{sec:discuss}

\subsection{Comparison with previous structure function determinations}
\label{subsec:compobs}

\begin{figure}
  \centering
  \includegraphics[width=\linewidth]{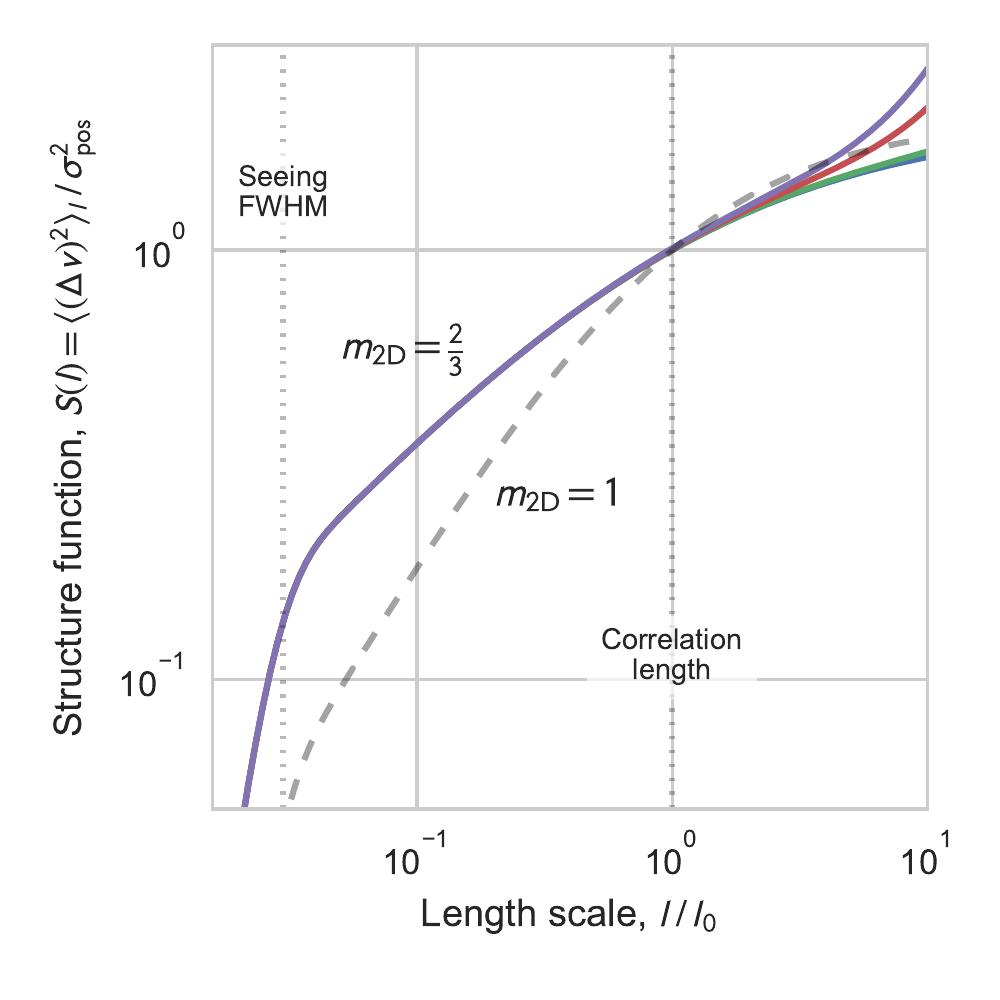}
  \caption{Idealized plane-of-sky normalized structure function
    \(S(l) = 2\left[ 1 - C(l) \right]\) for turbulent velocity
    autocovariance of the form
    \(C(l) = 1/\left[1 + (l/l_0)^{m_\mathrm{2D}}\right]\), where
    \(l_0\) is the correlation length of the turbulence. Results for
    Kolmogorov-like turbulence (\(m_\mathrm{2D} = 2/3\)) and a steeper
    spectrum (\(m_\mathrm{2D} = 1\)) are shown.  In the first case,
    different colored lines show the effect of an uncorrected
    linear velocity gradient on the scale of the map (size \(10
    l_0\)), with total amplitude of (top to bottom) 2.0, 1.0, 0.1, and
    0.0 times the turbulent velocity dispersion,
    \(\sigma_\mathrm{pos}\).  At small scales, the effect of seeing
    is shown, assuming a Gaussian FWHM of \(0.03 l_0\). 
  }
  \label{fig:sf-ideal}
\end{figure}

Figure~\ref{fig:sf-ideal} illustrates how various effects modify the
purely turbulent structure functions for a highly idealized case.  The
most promising range of length scales for measuring the turbulent
velocity spectrum is between a few times the seeing width and about
half the correlation length, \(l_0\), of the turbulence.  At scales
larger than \(l_0\), the structure function flattens as it tends
towards the asymptotic value of 2 for a homogeneous random field.  If
there is a linear velocity gradient across the map, then the structure
function will steepen again at the largest scales.  Alternatively, if
the turbulent velocity dispersion is inhomogeneous, being larger in
the center of the map than in the periphery, then the structure
function slope will become negative at the largest scales (not
illustrated).  The figure does not include the effects of noise, but
that is easily dealt with in the case that the noise is ``white''
(spatially uncorrelated, such as shot noise), since the effect is to
simply add a constant value to the structure function at all scales.
This can be estimated either from the variance of velocity differences
at separations significantly less than the seeing width or from
independent observations of the same spatial point, and then
subtracted from the observed structure function prior to analysis.
Spatially correlated ``noise'' due to systematic instrumental effects
is more difficult to deal with, since it will add a spurious
scale-dependent term to the structure function.

\begin{table}
  \newcommand\C[1]{\multicolumn{1}{c}{#1}}
  \caption{Comparison of structure function slopes}
  \label{tab:sf-lit}
  \centering
  \setlength{\tabcolsep}{0.45em}
  \begin{tabular}{lll@{}r@{\hspace{1.5em}}cl}\toprule
    &     &        & \C{\(\sigma^{2}\)} & \C{Range} & \\
    Reference & Ion & Method & \C{(\kmss{})} & \C{(\(''\))} & \C{Slope}\\
    \midrule
    \citealt{1992ApJ...387..229O}  & \oi{} & Mean & 3:\phantom{0} & 6--85 & \(0.68\):\\
    This paper & \sii{} & Mean & 5.4 & 7--32 & \(0.80 \pm 0.12\)\\
    This paper & \nii{} & Mean & 5.6 & 8--22 & \(0.82 \pm 0.11\)\\
    \citealt {1993ApJ...409..262W} & \siii{} & Comp A & 13.8 & 5--20 & \(0.92\):\\
    This paper & \ha{} & Mean & 9.4 & 8--22 & \(1.17 \pm 0.08\)\\
    This paper & \oiii{} & Mean & 10.2 & 8--22 & \(1.18 \pm 0.09\)\\
    \citealt{1988ApJS...67...93C}  & \oiii{} & Comp A & 13.7 & 3--15 & \(0.86 \pm 0.05\)\\
    This paper & \oiii{} & Comp A & 15.3 & 3--15 & \(0.73 \pm 0.05\) \\
    \bottomrule
  \end{tabular}
\end{table}

Previous studies of the velocity structure function in Orion have been
carried out based on slit spectra \citep{1988ApJS...67...93C,1992ApJ...387..229O,1993ApJ...409..262W}.  Table~\ref{tab:sf-lit} compares these results with our
own for different emission lines, ordered from lower to higher
ionization.  In spite of the differences in methodology, a broad
agreement is seen, with both the magnitude of the velocity dispersion
and the steepness of the structure function slope increasing with
ionization.  The most directly comparable methodology to our own is
that of \citealt{1992ApJ...387..229O}  who used the flux-weighted mean velocity
of the \oi{} \(\lambda\)6300 line.  The fitted range of \(6''\) to
\(85''\) extends to larger scales than in studies of other lines,
and this biases the slope determination towards lower values due to the
slight curvature of the structure function.    

The other studies are harder to compare with our own since they are
based on multi-component Gaussian fits to the line profiles, such as
shown in Figure~\ref{fig:gauss}.  In order to check the effects of
this methodological difference, we have calculated the \oiii{}
structure function for the strongest component of the three-Gaussian
decomposition of our line profiles (component~A, see
\S~\ref{sec:poslos}), with results that are also included
in Table~\ref{tab:sf-lit}.  We find a slightly larger plane-of-sky
velocity dispersion and shallower structure function than we found
using the mean velocity of the entire profile, which is consistent
with the results of \citet{1988ApJS...67...93C}  for the same line.
Experimentation shows that this is partly due to under-determination
of the Gaussian fits, particularly for components A and B, which are
severely blended at most positions.  This results in a fitting
degeneracy between the velocity separation
\(v_\mathrm{A} - v_\mathrm{B}\) and the flux ratio
\(F_\mathrm{A}/F_\mathrm{B}\) of the two components, which spuriously
contributes to the variation in \(v_\mathrm{A}\).  

The structure function slopes obtained by \citet{2016MNRAS.455.4057M}, based
on integral field spectroscopy with the MUSE instrument
\citep{Weilbacher:2015a} are significantly flatter than all other
studies, and we have decided not to include them in the comparison
table.  For example, they obtain a slope of \(0.29\) for \oiii{} and
\(0.0\) for [\ion{O}{1}].  This appears to be the result of
uncorrected fixed-pattern noise in their mean velocity maps, which can
be seen as tartan-like horizontal and vertical stripes in their
Fig.~10.  We will show in a following paper that once these
instrumental artifacts have been removed, the MUSE results are
consistent with other studies.

\subsection{Comparison with previous simulation results}
\label{subsec:compsim} 
In our previous paper \citepalias{{2014MNRAS.445.1797M}}, we used velocity
channel analysis and second-order structure functions to analyse our
3D numerical simulations of an evolving \ion{H}{ii} region in a turbulent
molecular cloud. The spectral index of the 3D power spectrum of the
underlying velocity fluctuations for our simulations was $n \sim
-3.1$, which is shallower than that predicted by Kolmogorov theory. We
found that for our velocity channel analysis of the emission lines of
the heavier ions, [\ion{O}{iii}], [\ion{N}{ii}] and [\ion{S}{ii}], the measured values of the
thin velocity slice spectral index $\gamma_\mathrm{t}$ agree to within
$0.1$ with the values derived using the formulae highlighted in
Table~\ref{tab:summary} using the simulation value of $n$. On the
other hand, the second-order structure functions we calculated from
our simulations proved less capable of reliably recovering the 3D
velocity power spectrum. These results motivated the work in the
current paper, that is, by applying velocity channel analysis to high
resolution spectroscopic observations we have the possibility of recovering the
spectral index of the underlying 3D velocity fluctuations.

In the present work, we do not have access to the underlying 3D
velocity distribution. Our VCA spectral indices show some similarities
but also some differences with respect to those obtained from the
simulations. We will discuss only the heavier ions, since the
H$\alpha$ is seriously affected by thermal broadening. To begin with,
for the simulations, we found that the [\ion{N}{ii}] and [\ion{S}{ii}] values of
$\gamma_\mathrm{T}$ (thick velocity slice spectral index) could be
considered ``shallow'' ($\gamma_\mathrm{T} > -3$), while the [\ion{O}{iii}]
had a ``steep'' spectral index ($\gamma_\mathrm{T} < -3$). For the
compensated power spectra $k^3P(k)$, a power-law index equal to the
critical value $\gamma_\mathrm{T} = -3$ would appear horizontal. We remark
that the 2D power spectra of the simulations are not at all noisy,
unlike the 1D power spectra presented in Section~\ref{subsec:vca} and,
moreover, a single power law is sufficient to cover most of the
wavenumber range. In contrast, our observationally derived power
spectra can be split into different regimes where different power laws
apply with sharp breaks between them (see Figs.~\ref{fig:HNOVCA} and \ref{fig:SIIVCA}). In the principal wavenumber range (regime~II in
Section~\ref{subsec:vca}), the spectral index of the thick velocity
slices, $\gamma_\mathrm{T}$, is  ``shallow'' for [\ion{N}{ii}] and [\ion{O}{iii}] but
apparently steep for [\ion{S}{ii}]. Thus,
while there is agreement in the [\ion{N}{ii}] spectral index between
observational and simulation results, the [\ion{O}{iii}] spectral index is
shallower in real life. The spectral index of [\ion{S}{ii}] is
misleading, since this emission line is affected by collisional
deexcitation for electron density greater than about $10^3$~cm$^{-3}$, which will suppress power at higher $k$ (smaller length
scales). This was not an issue in the numerical simulations because
the electron densities were lower than in the Orion Nebula. 

With regard to the power-law indices of the second-order structure functions, our simulations
found a clear sequence $1 > m_\mathrm{2D}(\mbox{OIII}) >
m_\mathrm{2D}(\mbox{H}\alpha) > m_\mathrm{2D}(\mbox{NII}) >
m_\mathrm{2D}(\mbox{SII})$. The observational results, summarized in
Table~\ref{tab:sfind}, show a similar sequence but the absolute values
are slightly different. We note that the offset of approximately 0.3 in $m_\mathrm{2D}$
values between simulation and observational results is no larger than
the spread in observationally derived results obtained with different
methodologies discussed in Section~\ref{subsec:compobs}.
The simulation structure function follows a clear power law over a
wide range of spatial separations (straight line in log-log space)
whereas the observational structure function has a slowly varying
power law as a function of separation scale.

Both the structure function power-law indices and the VCA
$\gamma_{\mathrm{t}} - \gamma_{\mathrm{T}}$ values are potential
diagnostics of the underlying velocity power spectrum. The differences
we find in these quantities between the simulations and the
observations are consistent with a steeper velocity spectrum for the
real nebula. On the other hand, the thick slice VCA spectral index
$\gamma_{\mathrm{T}}$, which is simply the spectral index of the
surface brightness power spectrum,  is
a diagnostic of the underlying emissivity power spectrum. The
simulation results give very different values for the [\ion{N}{ii}] and [\ion{O}{iii}]
thick-slice spectral indices, suggesting a very different spatial
distribution for these two emission regions. The observational values
are more similar for [\ion{N}{ii}] and [\ion{O}{iii}].

We remark here that in \citetalias{2014MNRAS.445.1797M}, we used
the  projected emission maps to estimate the mean radius of the nebula
in a given emission line and used this as the upper limit of the
separation range for the
power-law fit to the structure functions. The lower limit was set
to be 10 computational cells, i.e. the scale at which numerical
dissipation effects cease to be important. The corresponding
wavenumber ranges were used in the VCA fits. This is in contrast to
the present paper, where the VCA graphs clearly indicate the wavenumber
ranges of interest and the corresponding separation scales
are used to fit the structure-function power laws.

How can we explain the differences between the observational results
and those obtained from the numerical simulations? To begin with, the
simulations are not a model for the Orion Nebula: the
size of the simulation box was 4~pc, which, although similar in size to
the Extended Orion Nebula \citep{ODellHarris:2010}, is an
order of magnitude larger than the $3^\prime\times5^\prime$ observed
region studied in this paper. The simulation ran for 300,000~yrs, while
the Orion Nebula is of uncertain age ($< 2$~million years). The
simulations omitted several physical processes. The most important one
is probably the stellar wind from $\theta^1$~Ori~C
\citep{2009AJ...137..367O}, which will evacuate a cavity inside the
distribution of photoionized gas such that even the high ionization emission lines 
will come from a thick shell rather than a filled volume. This can
explain why the observed [\ion{O}{iii}] surface brightness spectral index is
shallower than in the simulations.
The wind bubble will also act as an obstacle around which the
photoionized gas streams away from the ionization front in a champagne
flow (see \citealp{2006ApJS..165..283A} for examples of
numerical simulations of this scenario). Furthermore, the limited
dynamical range of the simulations may have prevented the development
of a fully resolved turbulent energy cascade.

Finally, the real Orion Nebula contains a large population of
low-mass stars, some of which are sources of jets and Herbig-Haro
objects. Although the effect of such structures is spatially limited
and we have excised extreme velocity features from our sample used to
calculate the structure function (see Section~\ref{sssec:SOSF}), there still could be a small
contribution from bright features that lie more-or-less in the plane of the sky.

\subsection{Turbulent contribution to spectral line broadening}
\label{sec:linebroad}

So far we have concentrated on the slopes of the structure function
and power spectra, since they can be theoretically related to the
spectrum of underlying velocity fluctuations in the nebula.  However,
other questions require consideration of the magnitude of the velocity
fluctuations, as measured by different techniques.  Such questions
include whether turbulence alone is sufficient to account for the non-thermal
line broadening observed in the nebula, and whether that turbulence is
driven primarily by the large scale thermal expansion of the nebula,
or by smaller scale photoevaporation flows, and to what extent stellar
winds play a role. 

The plane-of-sky dispersion in centroid velocities
\(\sigma\pos(\bar{u})\) is the RMS width of the marginal distribution
along the \(\bar{u}\) axis of Figure~\ref{fig:observed-vmean-sigma},
which is the same as the quantity \(\sigma_{\mathrm{vc}}\) used to
normalize the structure functions in \S~\ref{sec:methods}.
Figure~\ref{fig:obs-sigma-sigma}(\textit{a}) summarizes the
observational determination of \(\sigma\los\) and \(\sigma\pos\),
while Figure~\ref{fig:obs-sigma-sigma}(\textit{b}) shows the same
quantities calculated for the turbulent \hii{} region simulation of
\citetalias{2014MNRAS.445.1797M}, as presented in Appendix~\ref{sec:appd}.  As
discussed above, the FWHM-derived and reflection-corrected
observational values are the most reliable, and these fall broadly in
the same region of the graph as do the evolutionary tracks, with
\(\sigma\pos \approx 3 \pm 1 \ \kms\) and \(\sigma\los \approx 6 \pm 1
\ \kms\).  

The fact that the line-of-sight velocity dispersion is roughly twice
the plane-of-sky velocity dispersion can be interpreted in at least
two different ways.  In the case of a homogeneous turbulent velocity
field with characteristic correlation length \(l_0\), the projection
from three to two dimensions over a line-of-sight depth \(H\) reduces
the plane-of-sky amplitude of fluctuations if \(l_0 < H\), a
phenomenon known as ``projection smearing'' \citep{{1951ZA.....30...17V},
  Scalo:1984a}.  Our \(l_0\) and \(\sigma\pos / \sigma\los\)
correspond to \(s_0\) and
\(\sigma_{\mathrm{obs}} / \sigma_{\mathrm{true}}\) in Fig.~1 of
\citet{Scalo:1984a}, from where it can be seen that a value of \(0.5\)
requires \(l_0 / H \approx 0.02\)--\(0.1\), depending on the steepness
of the velocity fluctuation spectrum.  The results from our structure
function analysis (\S~\ref{subsec:sf}) imply a correlation length
\(l_0 \approx 0.1\)--\(0.2\)~pc for all lines, which would require a
very large line-of-sight depth \(H > 1\)~pc in order to explain the
observed \(\sigma\pos / \sigma\los\) by projection smoothing.  This is
inconsistent with independent evidence \citep{Baldwin:1991a,
  {2001ARAA..39...99O}, {2007AJ....133..952G}} that the emitting layer thickness is
much smaller than this in the region covered by our maps:
\(H \approx 0.01\)--\(0.3\)~pc, being thinner on the West side and for
the lower ionization lines.

The same discrepancy arises when the projection smearing argument is
applied to our simulated \hii{} region.  In this case,
\(l_0 \approx 0.5\)--\(1\)~pc for evolutionary times later than
\(0.15\)~Myr (Fig.~A4 of \citetalias{2014MNRAS.445.1797M} ), thus requiring
\(H > 5\)~pc, which is clearly unacceptable since it is bigger than the
computational box of the simulation.  For both the simulation and the
observations, it is clear that \(l_0 \approx H\) for the
high-ionization lines and \(l_0 > H\) for the low-ionization lines.
Therefore, projection smearing of the large-scale fluctuations is
negligible and cannot explain the difference between \(\sigma\pos\)
and \(\sigma\los\). 

\begin{figure}
  \centering
  \includegraphics[width=\linewidth]{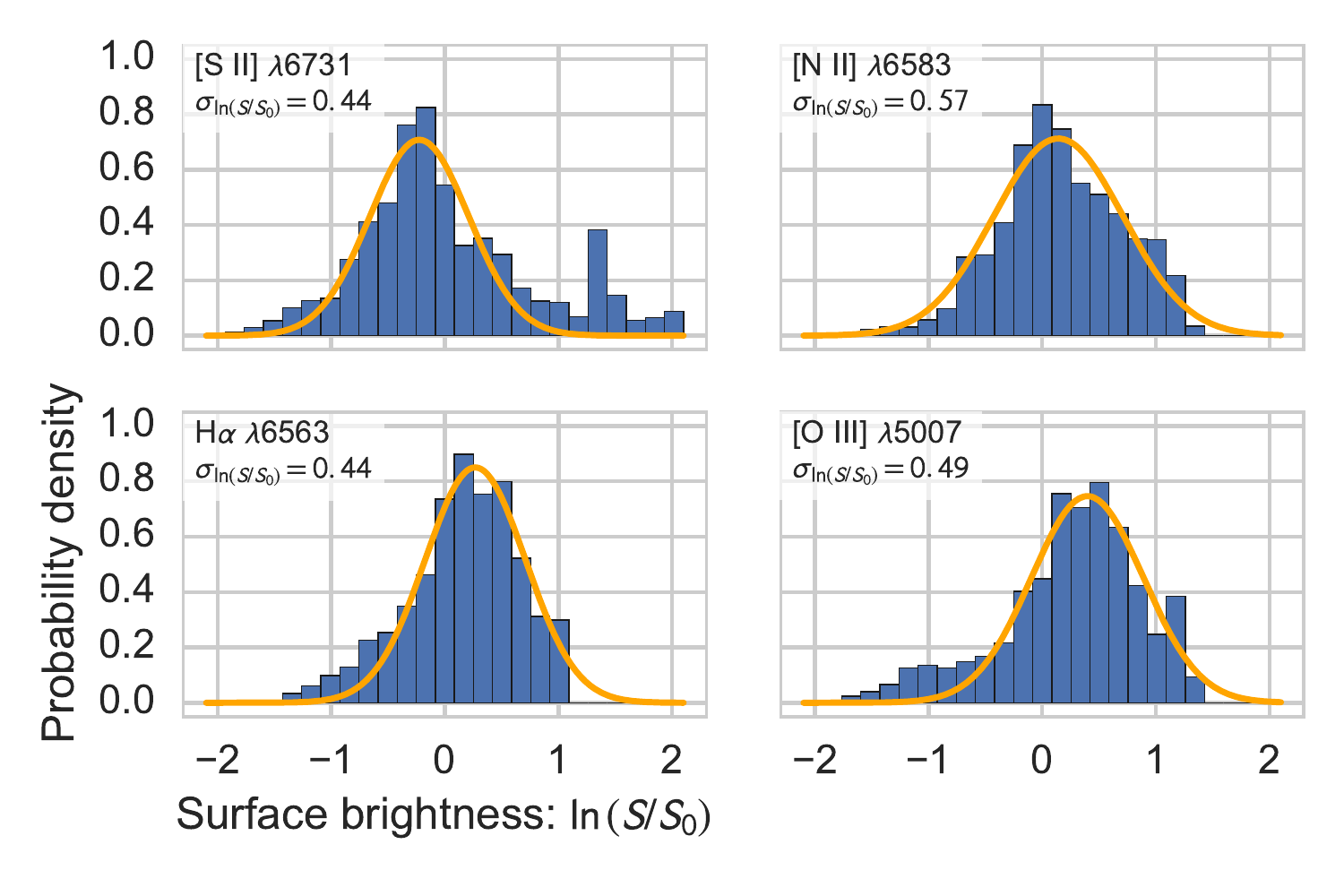}
  \caption{Probability density functions (PDF, blue histograms) of
    observed surface brightness maps in different emission lines after
    binning \(16 \times 16\) pixels to eliminate small scale
    variations.  The surface brightness is normalized by the mean
    value and shown on a natural logarithmic scale, and the PDFs are
    weighted by the surface brightness itself.  A log-normal
    distribution (orange lines) is fitted to the core of each PDF
    (only the part of the histogram with probability density \(> 0.3\)
    times the maximum).  The fitted RMS width \(\sigma\) of the
    log-normal distribution is indicated on each panel.}
  \label{fig:surf-bright-pdf}
\end{figure}

\newcommand\Efrac{\ensuremath{_{\scriptscriptstyle E/E_0}}}
\newcommand\lnSfrac{\ensuremath{_{\scriptscriptstyle \ln S/S_0}}}
\newcommand\Sfrac{\ensuremath{_{\scriptscriptstyle S/S_0}}}
\newcommand\denfrac{\ensuremath{_{\scriptscriptstyle \rho/\rho_0}}}

A second, contrasting interpretation of the evidence would be in terms
of large-scale, ordered motions.  Consider an emission shell that
expands at velocity \(v\).  If the shell emissivity is homogeneous,
then the integrated line profile is rectangular, with mean velocity
\(\bar{u} = 0\) and velocity width \(\sigma\los = v/2\).  Furthermore,
the spatially resolved line profile at any point will also have
\(\bar{u} = 0\), so that \(\sigma\pos = 0\).  However, if there
are emissivity fluctuations between different parts of the shell, then
\(\bar{u}\) will fluctuate on the plane of the sky, according to
the relative brightness of the red-shifted and blue-shifted
hemispheres.  
The required RMS fractional variation in the emissivity on the scale
of the shell diameter is found to be \(\sigma\Efrac \approx
\sigma\pos/\sigma\los\). 

The observed large-scale (\(> 9''\)) brightness fluctuations are
illustrated in Figure~\ref{fig:surf-bright-pdf}, which shows
log-normal fits to the PDFs of surface brightness, \(S\), after
normalizing by the mean, \(S_0\), and binning the maps at
\(16 \times 16\) pixels.  The RMS width of the log-normal PDF,
\(\sigma\lnSfrac\) is seen to be in the range 0.45--0.6 for all lines.
This is related to the RMS fractional brightness fluctuation as
\(\sigma^2\lnSfrac = \ln(1 + \sigma^2\Sfrac)\), or
\(\sigma\lnSfrac \approx \sigma\Sfrac\) if \(\sigma\Sfrac < 1\).  The
relationship between \(\sigma\Efrac\) and \(\sigma\Sfrac\) depends on
both line-of-sight projection \citep{Brunt:2010b}, which tends to make
\(\sigma\Sfrac < \sigma\Efrac\), and fluctuations in the foreground
dust extinction, which have the opposite effect of increasing
\(\sigma\Sfrac\).  
The first effect dominates, so
 that the surface brightness PDFs imply
 \(\sigma\Efrac \ga 2 \sigma\Sfrac \sim 1\) (see \S~\ref{sec:does-veloc-turb} for details). 
This is larger than  
the value derived in the previous paragraph,
which implies that emissivity
fluctuations combined with an ordered velocity field are entirely
sufficient to explain the observed plane-of-sky variation in mean
velocities, without requiring any fluctuations in the velocity field
itself.

\newcommand\champ{\ensuremath{_{\mathrm{cham}}}}
\newcommand\turb{\ensuremath{_{\mathrm{turb}}}}
Although it is a priori unlikely that there are \textit{no} velocity
fluctuations in the ionized gas, this is yet another reason why the
structure function of the mean velocity is not an effective diagnostic
of these fluctuations in the presence of strongly inhomogeneous
emissivity and large-scale velocity gradients.  In Orion, the ordered
large scale expansion of the nebula is an asymmetrical champagne flow
away from the background molecular cloud \citep{1973ApJ...183..863Z}, which
produces systematically larger blue-shifts with increasing ionization
(e.g., Fig.~11 of \citealp{Baldwin:2000a}).  This offers a simple
method for estimating the relative contribution of ordered versus
turbulent motions to the total velocity dispersion.  The mean
systematic difference between the \oi{} and \oiii{} centroid
velocities is \(\delta u = 9.4~\kms\) (Table~2 of
\citealp{2008RMxAA..44..181G}), which gives a champagne-flow
contribution to the velocity dispersion of \(\sigma\champ \approx 0.5\,
\delta u = 4.7~\kms\).   The turbulent contribution to the velocity
dispersion is then \(\sigma\turb \approx (\sigma\los^2 -
\sigma\champ^2)^{1/2} = 3.7~\kms\).  The uncertainties in this
analysis are large, so that all that can be confidently asserted is
that the ordered and turbulent velocity dispersions are roughly equal
with \(\sigma\champ \approx \sigma\turb = 4\)--\(5~\kms\).

\subsection{What is the significance of the 22 arcsec and 8 arcsec
  length scales?}
\label{sec:what-significance-22}

The velocity channel analysis suggests that two length scales are
important in the Orion Nebula, corresponding to the limits of
regime~II (see Fig.~5). The break in power law at 22~arcsec and
8~arcsec occurs for both thin and thick velocity slices. This
indicates that it is a feature of the emissivity power spectrum, and
not the velocity power spectrum. Below 8~arcsec (regime~III), the
emissivity power spectrum is very steep in all the lines, indicating
that small-scale fluctuations are relatively unimportant.  Above
22~arcsec (regime~I), the power spectrum \(P(k)\) is very flat,
similar to the noise-dominated spectrum in regime~IV, suggesting that
fluctuations are relatively uncorrelated on larger scales.  This is
underlined by the analysis in Appendix~\ref{sec:toy-model-surface},
where it is shown that a combination of Gaussian brightness peaks with
widths (FWHM) from \(\approx 4''\) and \(12''\) can capture the broad
features of the observed power spectra.\footnote{The factor of two
  difference between these widths and the length scales discussed
  earlier in the paragraph is because a fluctuation consists of both
  a peak and a trough, and so the width of the peak is only one half
  of the fluctuation wavelength.}

The outer scale of \(22''\) coincides with scale where the structure
functions reach a value of unity (Fig.~8), which corresponds to the
correlation length, \(l_0\), of the velocity fluctuations (see
Fig.~14).  It is therefore plausible to associate this scale, which
corresponds to a physical size of \(\approx 0.05\)~parsec, with the
driving scale of turbulence in the nebula.  The inner scale of
\(8''\), corresponding to a physical size of \(\approx 0.02\)~parsec,
is harder to associate with any particular process since the structure
functions (Fig.~8) show no apparent feature at this scale.

\subsection{Does velocity turbulence cause the surface brightness
  fluctuations?}
\label{sec:does-veloc-turb}

The surface brightness fluctuations on the plane of the sky are
primarily caused by emissivity fluctuations within the nebular volume,
which are in turn caused by fluctuations in electron density,
temperature, and ionization.  The temperature and ionization
dependence of the emissivity is very different for each line, but the
electron density dependence is similar in all cases, being
\(\propto N_e^2\) in the low density limit, which is appropriate for
all but the \sii{} lines.  It therefore seems likely that any
\emph{commonalities} in the statistics between all the different
emission lines will give us information about the electron density
fluctuations within the nebula.

In \S~\ref{sec:linebroad} it was shown that the rms fractional
surface brightness variation in 2D is \(\sigma\Sfrac \approx 0.5\) for
all lines, and the rms emissivity variation in 3D is predicted to be
\(\sigma\Efrac = \xi \sigma\Sfrac\), where the ``de-projection
factor'' is \(\xi = 2\)--\(3\) \citep{Brunt:2010a}.  On the other
hand, if the emissivity fluctuations are due to variations in the
density squared, then the rms fractional 3D density variation is
\(\sigma\denfrac = 0.5\, \sigma\Efrac\), which approximately cancels
out the de-projection factor so that
\(\sigma\denfrac \approx \sigma\Sfrac \approx 0.5\).  If the density
fluctuations are \emph{caused} by the turbulent velocity fluctuations,
then numerical simulations \citep{Konstandin:2012a} show that there is
a linear relationship between \(\sigma\denfrac\) and the rms Mach
number, \(M\), of the turbulence: \(\sigma\denfrac = b M\), where
\(b = 1/3\) to \(b = 1\), depending on whether the turbulent driving
is primarily solenoidal or compressive.  The rms Mach number is the
ratio of the velocity dispersion to the ionized isothermal sound speed
\(M = \sigma_u / c_i\), where 
\(\sigma_u = \sigma\turb \approx 4~\kms\)
(see \S~\ref{sec:poslos}) and \(c_i \approx 11~\kms\).  
Thus, \(M \approx 0.36\)
so that, given \(b<1\), an upper limit to the turbulent contribution to the
density fluctuations is \(\sigma\denfrac \approx 0.36\).

Furthermore, the slopes of the surface
brightness fluctuation spectra in regime~II are significantly
shallower than expected from a top-down turbulent cascade in the
subsonic limit \citep{Konstandin:2015a}.  A similar result is seen in
our numerical simulations (see \S~\ref{subsec:compsim}), with the important difference
that in the simulation the velocity spectrum is also shallow, whereas
the thin-slice VCA analysis for the observations (\S~\ref{sssec:VCAR2}) is
consistent with a Kolmogorov slope for the velocity fluctuation
spectrum in regime~II. 

We therefore require a further mechanism to explain the roughly 50\%
of the variance in ionized density that cannot be accounted for by
turbulent velocity fluctuations.  This could plausibly be provided by
the bright-rimmed structure of the photoevaporation flows away from
dense molecular globules and filaments (e.g., \citep{Bertoldi:1990a,
  {2009MNRAS.398..157H}}), which are responsible for driving the turbulence.
We have calculated the emissivity-weighted density PDF for a simple
model of a single spherically divergent, isothermal evaporation flow
from a D-critical ionization front \citep{Dyson:1968a} and find
\(\sigma\denfrac = 0.56\).  For an an ensemble of such flows with
varying peak densities the \(\sigma\denfrac\) would be even higher, so
that in order for their global contribution to rival that of the
velocity fluctuations it is sufficient that a fraction
\(0.1\)--\(0.5\) of the total emission should come from such flows.

\section{Speculation}
\label{sec:speculation}

\begin{figure}
  \centering
  \includegraphics[width=\linewidth]{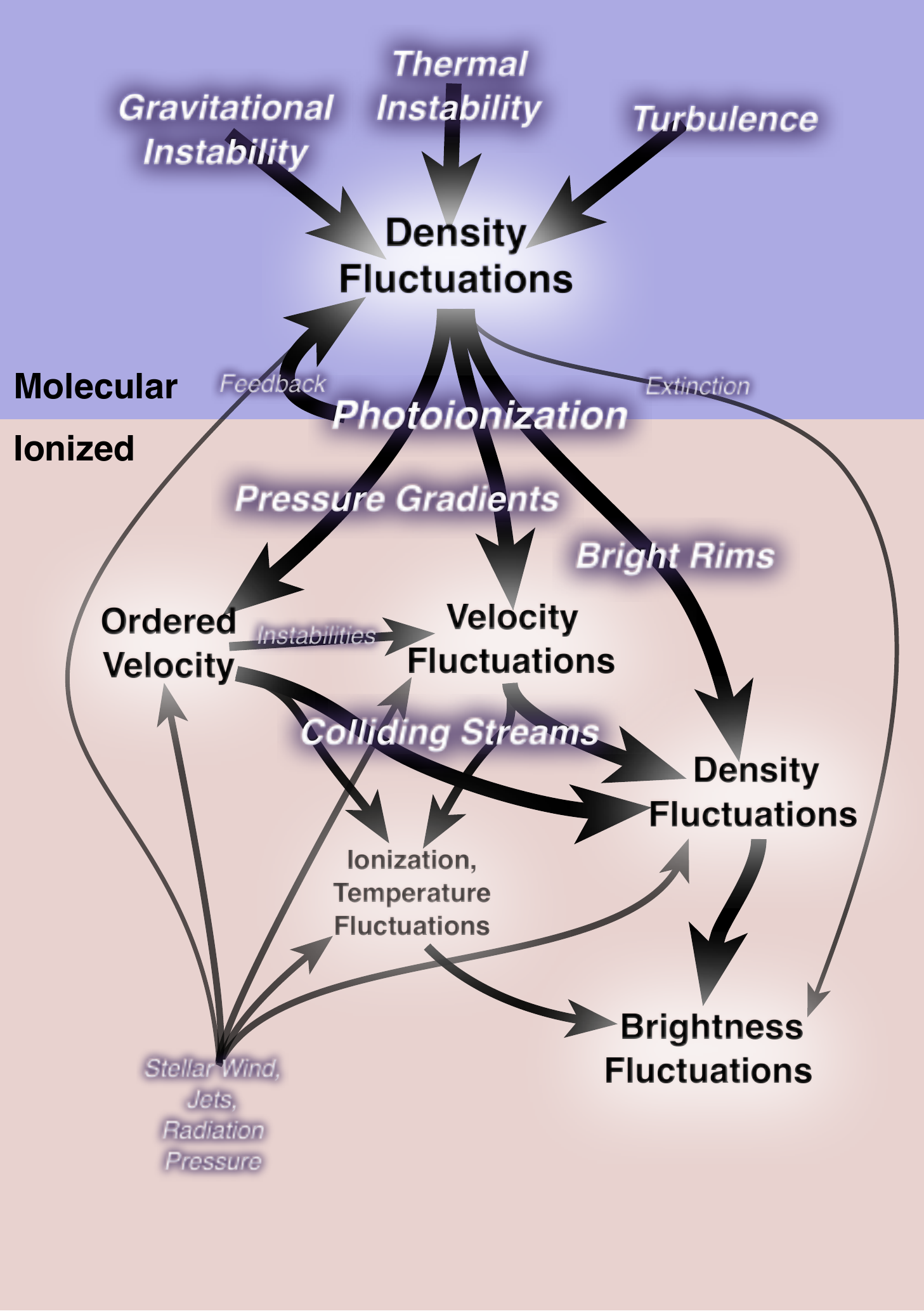}
  \caption{Causal relationships (arrows) between different types of
    fluctuations (black text) in molecular clouds (above) and \hii{}
    regions (below) via different physical processes (white text).
    Line thickness and text size is proportional to the relative
    importance of each process in the Orion Nebula.}
  \label{fig:causality-flow}
\end{figure}

We offer a speculative account of the complex web of physical
processes that give rise to to the velocity and brightness
fluctuations that we observe in the Orion Nebula.  This is illustrated
in Figure~\ref{fig:causality-flow}, where the most important causal
links are shown by thick arrows and secondary processes by thin
arrows.  

The principal origin of all structure in the \hii{} region is
the highly filamentary and clumpy density structure in the molecular
cloud from which it is emerging, which in turn has its origin in some
combination of thermal and gravitational instability and supersonic
turbulence \citep{Padoan:2002a, Ballesteros:2011a}.  In the molecular
gas, thermal pressure is negligible compared with magnetic pressure,
turbulent ram pressure, and the gravitational potential.  However, the
large temperature increase that accompanies photoionization means that
thermal pressure dominates in the \hii{} region, so that density
gradients are converted into pressure gradients that can accelerate
the gas.  The fractal nature of the molecular density means that gas
acceleration occurs on multiple scales, from the global outward radial
expansion of the \hii{} region (which in Orion is a highly one-sided
champagne flow) down to photoevaporation flows from individual
globules.   One piece of evidence for a direct connection between
molecular density fluctuations and ionized velocity fluctuations is
that \citet{Kainulainen:2016a} find correlation lengths of order \(0.08\)~pc
for the separations of molecular cores along the ridge that lies
behind the Orion Nebula, which is similar to the correlation lengths
we find for the velocity fluctuations in the nebula. 

Ionized density fluctuations can arise directly from the molecular
density fluctuations, such as the bright rims at the edges of
photoionized globules \citep{2009MNRAS.398..157H}, and this is most important
in the lower ionization zones near the ionization front where the
\sii{} and \nii{} emission is strong.  In the more highly ionized
interior of the nebula, it is collisions between opposing velocity
streams that produce the ionized density fluctuations, but these
fluctuations are less extreme than those seen in molecular gas because
the turbulence is subsonic.

The ionized density fluctuations are the primary determinant of the
emission line surface brightness fluctuations (\S~\ref{sec:does-veloc-turb}), although
ionization and temperature structure can make a contribution for
particular lines and there is also a direct contribution from
foreground molecular density fluctuations via dust extinction
\citep{ODell:2000a}.

Finally, a variety of other processes, such as O~star winds, radiation
pressure, and bipolar jets from young stars can play a secondary role
in stirring up gas motions.  In the case of the Orion Nebula, evidence
for the influence of stellar wind interactions is restricted to the
central \(0.05\)~pc \citep{Garcia-Arredondo:2001a} and the low-density
western outskirts \citep{Gudel:2008a}, and they seem to have little
influence on the bulk of the nebular gas.  Stellar wind effects are
more important in older and more massive regions that contain LBV and
Wolf-Rayet stars (e.g., \citealp{Smith:2007a}).  Similarly, radiation
pressure, although unimportant in Orion, becomes much more important
in higher luminosity regions \citep{Krumholz:2009a}.  Herbig-Haro jets
and bowshocks dominate the far wings (\(\delta u \sim 50~\kms\)) of
the velocity distribution in Orion \citep{Henney:2007b}, but the total
kinetic energy of these high velocity flows is relatively low, so that
the effect on the global velocity statistics is minor.

\section{Summary}
\label{sec:summary-conclusions}

We have used statistical analysis of high-resolution spectroscopic
observations of optical emission lines in the central
\(0.4 \times 0.6\)~pc ($3^\prime\times 5^\prime$) of the Orion Nebula in order to characterize the
turbulence in the ionized gas. The analysis has been guided and
informed by radiation hydrodynamic simulations of \hii{} region
evolution. The techniques that we have applied are:
\begin{enumerate}[(1.)]
\item Second-order structure function of velocity centroids (\S~\ref{sssec:SOSF}), which
  gives the variation as a function of plane-of-sky separation of the
  differences in average line-of-sight velocity.
\item Velocity channel analysis (VCA; \S~\ref{subsec:MVCA}), which compares the spatial
  power spectrum slope of velocity-resolved and velocity-integrated
  emission profiles of the same line.
\item Line-width analysis (\S~\ref{sec:poslos}), which is sensitive to velocity fluctuations
  along the line of sight
\item Probability density function (PDF; \S~\ref{sec:linebroad}) of the surface brightness in
  different lines
\end{enumerate}
Our principal empirical findings are as follows:
\begin{enumerate}
\item The VCA technique is the most reliable means of determining the
  spectrum of velocity fluctuations in the ionized gas (\S~\ref{sssec:VCAR2}), and we find
  consistent evidence from both low and high ionization lines for a
  Kolmogorov-type spectrum (\(\delta u \sim l^{1/3}\)) for length
  scales, \(l\), between \(0.05\)~pc (\(\approx 22''\)) and \(0.02\)~pc
  (\(\approx 8''\)).  Unfortunately, VCA can not be applied if the
  thermal or instrumental line width is larger than the velocity
  differences of interest (Appendix~\ref{sec:thin}), which rules out its application to the
  \ha{} line and to scales smaller than \(0.02\)~pc. 
\item The structure functions show systematic trends with degree of
  ionization (\S~\ref{sssec:SFPI}).  Higher ionization lines tend to show larger
  autocorrelation scales, larger total plane-of-sky velocity
  dispersions, and steeper slopes than lower ionization lines.  The
  changes in slopes are difficult to interpret because of the
  influence of projection smearing and sensitivity to details of the
  observational methodology (\S~\ref{subsec:compobs}).
\item The characteristic length of \(0.05\)~pc is special in at least
  two ways, corresponding to both the autocorrelation scale of
  velocity differences for low-ionization lines (Figs.~\ref{fig:HNOSF}, \ref{fig:SIIVCA}, \ref{fig:sf-ideal}) and also a break in
  the power spectrum of surface brightness fluctuations in all lines (Figs.~\ref{fig:HNOVCA}--\ref{fig:vcahoriz}).
  We suggest that this is the dominant scale for density fluctuations
  in the nebula (\S~\ref{sec:what-significance-22}) and is also the main driving scale of the
  turbulence. A further break in the surface brightness power spectra
  occurs at the smaller scale of \(0.02\)~pc (\(\approx 8''\)), but
  there is no obvious feature in the structure functions at this
  scale.
\item Comparison of the application of turbulent diagnostics to numerical simulations \citepalias{2014MNRAS.445.1797M} with application of the same diagnostics to Orion leads us to conclude (\S~\ref{subsec:compsim})  that even the high-ionization line emission (eg [OIII]) is confined to a thick shell and does not fill the interior of the nebula. Furthermore, the underlying power spectrum is shallower in the simulations, implying that small-scale turbulent driving is less important in the nebula than it is in the simulations.

\item There are three lines of evidence suggesting that the velocity
  fluctuations are not homogeneous on the largest scales, but rather
  that the turbulent conditions themselves vary, both across the sky and
  along the line of sight, on scales larger than the velocity
  autocorrelation length of \(0.05\)--\(0.15\)~pc:
  \begin{enumerate}
  \item The structure function slope of the \nii{} line is
    significantly steeper in the southern half of our observed field
    than in the northern half (Fig.~\ref{fig:HNOSF}).
  \item The plane-of-sky velocity dispersion $\sigma_\mathrm{vc}$ increases with increasing
    ionization (Table~\ref{tab:sf-lit}), implying an increasing amplitude of fluctuations
    towards the interior of the nebula
  \item The line-of-sight non-thermal velocity dispersion (after
    removing the confounding effect of dust scattering; \S~\ref{sec:poslos}) is typically
    twice as large (\(\approx 6~\kms\)) as the plane-of-sky velocity
    dispersion (\(\approx 3~\kms\)).  In order to explain this ratio
    in terms of a homogeneous turbulent layer, the line-of-sight depth
    of the layer would need to be at least 10 times the velocity
    autocorrelation length, which is unrealistically large (\S~\ref{sec:linebroad}).  Instead,
    the result is more naturally explained by large-scale velocity
    gradients (such as radial expansion), combined with emissivity
    fluctuations along the line of sight.
  \end{enumerate}
\item The turbulent and ordered 
  components of the velocity dispersion ($\sigma\turb$ and
  $\sigma\champ$, respectively) are of similar magnitude: 
  estimated to be \(4 \mbox{--}
  5~\kms\)
  (\S~\ref{sec:linebroad}). 
\item The PDF of surface brightness fluctuations is approximately
  log-normal with a fractional width of \(0.4\)--\(0.5\) in all lines (Fig.~\ref{sec:linebroad}).
Turbulent velocity fluctuations can only account for half of the
variance in surface brightness. The remaining part may be due to
density gradients in photoevaporation flows (\S~\ref{sec:does-veloc-turb}).
\end{enumerate}

\section*{Acknowledgements}

SNXM acknowledges IMPRS for a PhD research scholarship. SJA, WJH and SNXM would like to thank DGAPA-UNAM for
financial support through projects PAPIIT IN101713 and IN112816. 
This work has made use of NASA's Astrophysics Data System.

\appendix

\section[]{Positions of high-resolution [S\,{\sevensize II}] observations}
\label{sec:appa}
\begin{figure}
\centering
\includegraphics[width=\columnwidth]{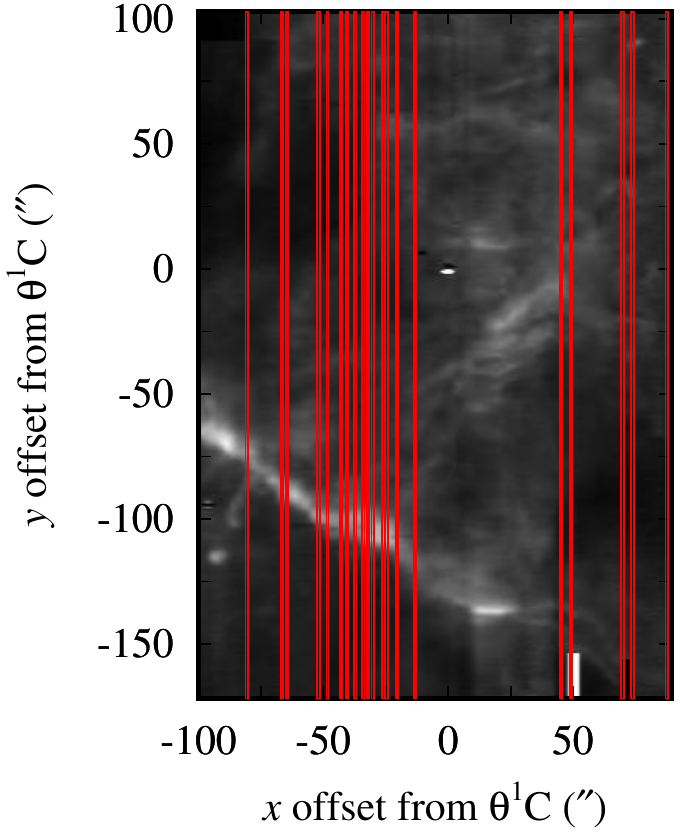}
\caption{Positions of the 20 high spatial resolution [\ion{S}{ii}] slits superimposed on the [\ion{S}{ii}]\,$\lambda$6716
  surface brightness image.}
\label{fig:siislits}
\end{figure}
 The [\ion{S}{ii}]~6716, 6731\AA\ observations consist of 92 North-South
pointings, 37 observed at KPNO with the same characteristics as the
H$\alpha$, [\ion{N}{ii}] and [\ion{O}{iii}] observations used in this paper and 55
observed at OAN-SPM. Of this latter group, 20 pointings have a high
velocity resolution of 6~km~s$^{-1}$ and the remaining 35 pointings
have a lower velocity resolution of 12~km~s$^{-1}$. All 92 datasets
are used to construct the velocity moment maps shown in
Figure~\ref{fig:masksii}. However, only the 20 highest resolution
observations are useful for the VCA calculations, since the other
observations are too affected by noise at high wavenumber. The
positions of the longslits for these 20 observations are shown in Figure~\ref{fig:siislits}

\section[]{Positions of [O\,{\sevensize III}] horizontal slits}
\label{sec:appb}
\begin{figure}
\centering
\includegraphics[width=\columnwidth]{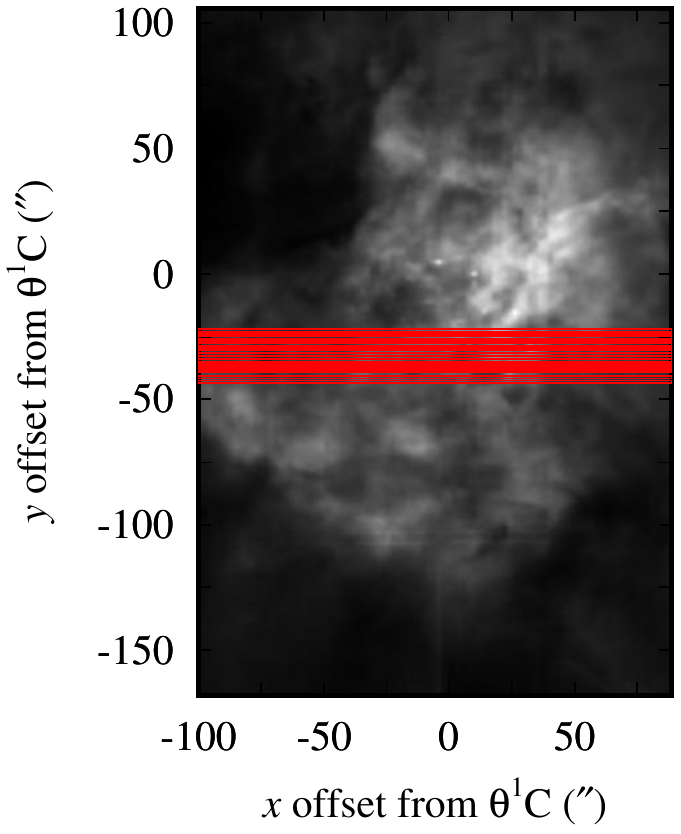}
\caption{Positions of the horizontal slits superimposed on the [\ion{O}{iii}]
  surface brightness image.}
\label{fig:ohoriz}
\end{figure}
 The main dataset consists of North-South oriented longslit
spectra. We can test whether orientation affects our velocity channel
analysis by examining a supplementary data set for the [\ion{O}{iii}] 5007
emission line consisting of observations perpendicular to the
main data set (i.e., oriented East-West) made at OAN-SPM. Fifteen spectra were obtained in steps of
$1.4^{\prime\prime}$ starting at $23^{\prime\prime}$ south of
$\theta^1$Ori~C and proceeding south. The slit positions and
orientations are indicated in Figure~\ref{fig:ohoriz} and the
observations are described in \citet{2015AJ....150..108O}. 

\section{Are the thin slices really thin?}
\label{sec:thin}

Our analysis of the power-law indices in Section~\ref{sec:analysis}
relates the difference between the spectral index obtained from the
thin slices and that of the thickest velocity slice to the underlying
3D power-spectrum index of the velocity fluctuations. We based our
choice of thin slice ($\delta v = 4$~km~s$^{-1}$) on the thinnest
slice not affected by instrumental broadening or thermal broadening. \citet{2003MNRAS.342..325E} discuss the criteria for a velocity channel to be thin or thick, which depends not only on the number of channels into which the total velocity range is divided, but also on the scale. For $N$ channels spanning the velocity range $\Delta v = v_\mathrm{max} - v_\mathrm{min}$ a channel $\delta v \equiv \Delta v/N$ is ``just thin enough'' if
\begin{equation}
N > \frac{\Delta v}{\sigma_\mathrm{L}}\left(\frac{L}{r}\right)^{m/2}
\label{eq:thincrit}
\end{equation}
where $L$ is a reference scale, $\sigma^2_\mathrm{L}$ is the
dispersion of velocities of points separated by this distance, $r$ is
the scale of interest and $m$ is the spectral index of the underlying
velocity fluctuations (3D structure function). The problem here is
that both $\sigma_\mathrm{L}$ and $m$ refer to the 3D velocity
distribution, while we have access to only the line-of-sight
velocities and can calculate only the 2D structure function of the
velocity centroids. We do, however, also have the line-of-sight
velocity linewidths and, as discussed in Section~\ref{sec:poslos} we
can estimate the turbulent linewidth $\sigma_\mathrm{los}$ (actually,
we use $\langle \sigma_{\mathrm{los}}\rangle_{\mathrm{pos}}$, the mean
line-of-sight velocity dispersion shown in Figure~\ref{fig:obs-sigma-sigma}). Assuming
that the line-of-sight turbulent linewidth $\sigma_\mathrm{los}$ is related to the
velocity dispersion at the largest scales before there is
decorrelation, then we have $\sigma_\mathrm{L} = \sqrt{2}
\sigma_\mathrm{los}$ and the scale $L$ corresponds to the local maximum in the second-order structure function, i.e the turnover point. For the spectral index, $m$, we use the Kolmogorov theoretical value $m_\mathrm{3D} = 2/3$, which we also derive from the VCA.

In particular, for the [\ion{O}{iii}]\,$\lambda$5007 emission line we obtain $\sigma_\mathrm{L} \simeq 1.4\times 6$~km~s$^{-1}$ and $L \simeq 86$~arcsec. Putting everything together, we can rearrange Equation~\ref{eq:thincrit} to find the minimum scale for which our $\delta v = 4$~km~s$^{-1}$ velocity slices are indeed thin:
\begin{equation}
r > \left(\frac{\delta v}{\sigma_\mathrm{L}}\right)^{2/m} L
\label{eq:thinscale}
\end{equation}
giving $r > 9$~arcsec. This is consistent with our VCA (see Fig.~\ref{fig:HNOVCA}), where the spectral indices of the thin velocity slices cease to make sense (they become steeper than the thick slice slopes) for scales smaller than $\sim 8$~arcsec. For the range corresponding to regime~II of the VCA, our thin velocity slices with $\delta v ~ 4$~km~s$^{-1}$ are in fact ``thin enough'' for the [\ion{O}{iii}] emission line.

For the [\ion{S}{ii}] emission lines, the value of $\sigma_\mathrm{turb} = 7$~km~s$^{-1}$ and the turnover in the structure function occurs at $L \sim 104$~arcsec and we obtain a minimum scale $r > 7$~arcsec, which again is consistent with our VCA (see Fig.~\ref{fig:SIIVCA}). Finally, for the [\ion{N}{ii}]\,$\lambda$6583 emission line, we find $\sigma_\mathrm{turb} = 6.5$~km~s$^{-1}$ and the turnover scale is $L\sim 107$~arcsec, from which we obtain $r > 9$~arcsec, which again is entirely consistent with our velocity channel analysis (see Fig.~\ref{fig:HNOVCA}).

\section[]{\boldmath Plane-of-sky versus line-of-sight variations in velocity from
simulated H\,{\sevensize II} region}
\label{sec:appd}
\begin{figure}
  \centering
  \setkeys{Gin}{width=0.47\linewidth}
  \begin{tabular}{@{}ll@{}}
    (\textit{a}) & (\textit{b}) \\
    \includegraphics{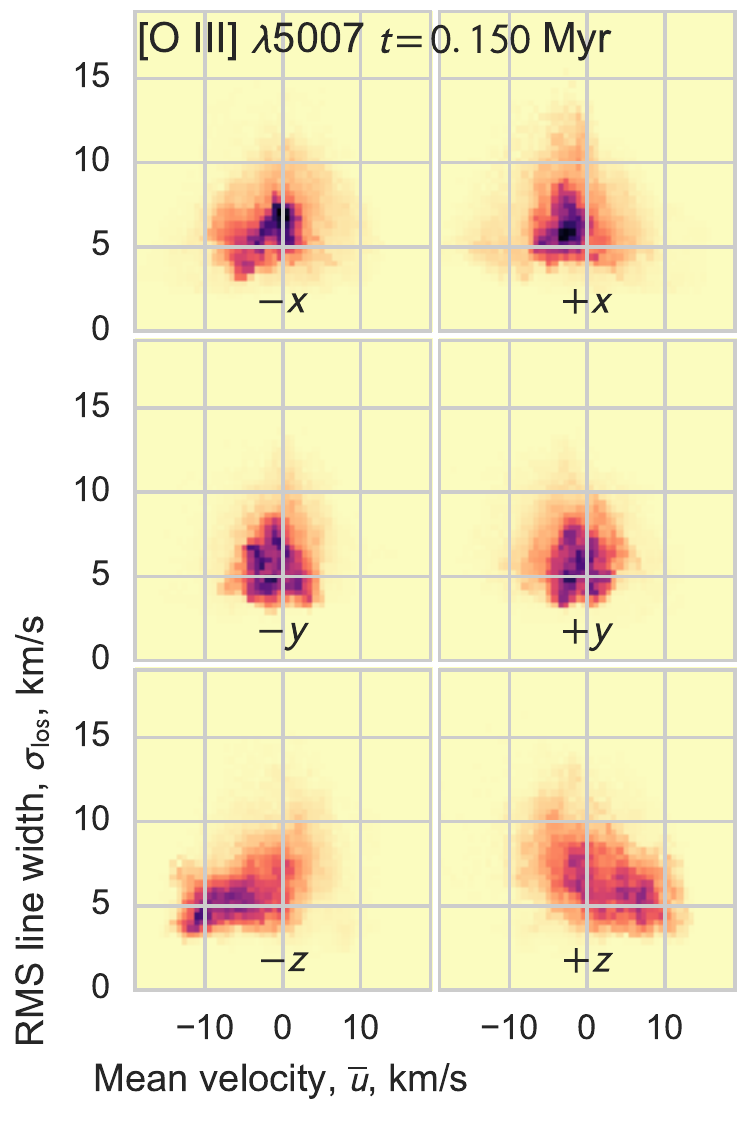}
    & \includegraphics{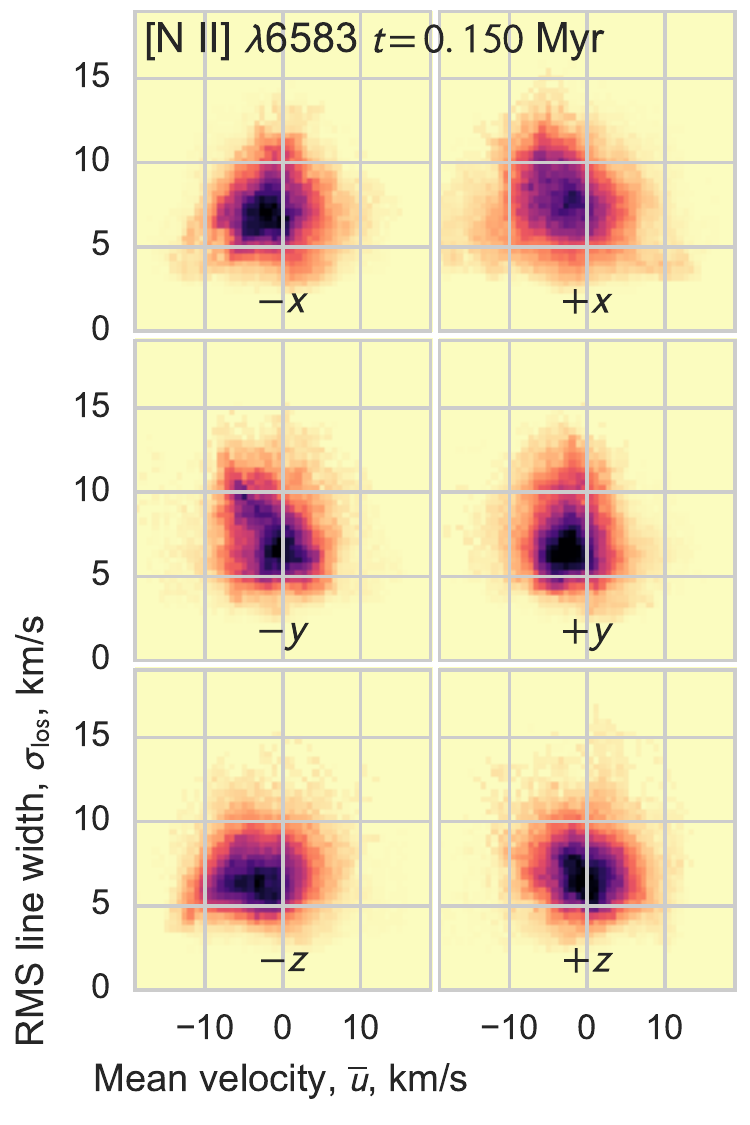}\\
    \\
    (\textit{c}) & (\textit{d}) \\
    \includegraphics{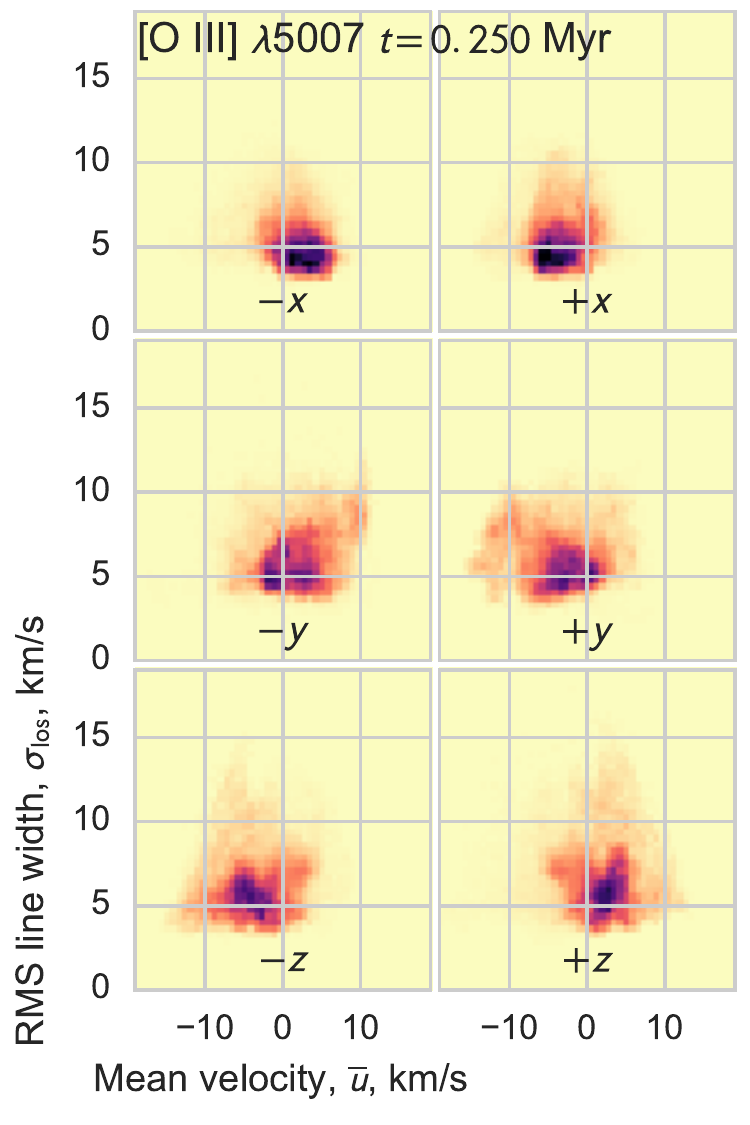}
    & \includegraphics{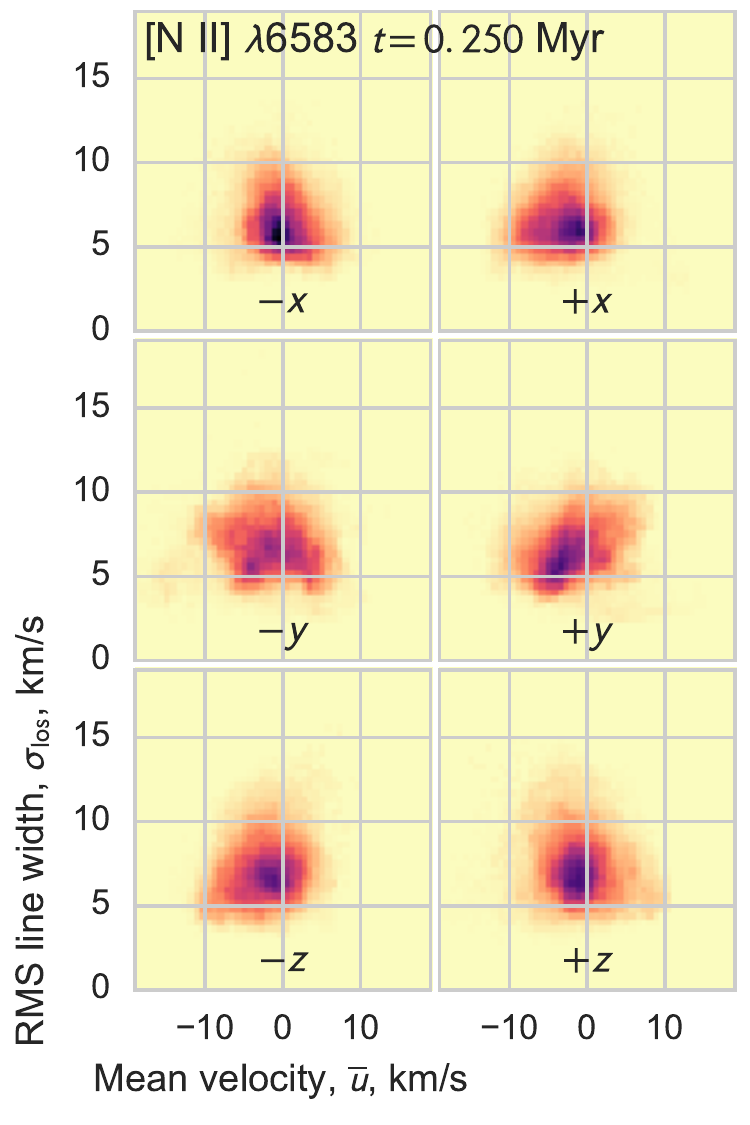}\\
  \end{tabular}
  \caption{Joint probability densities of mean velocity \(\bar{u}\)
    and line-of-sight velocity dispersion \(\sigma\los\) for two
    representative times in the evolution of a simulated \hii{}
    region: 0.15~Myr (\textit{a} and \textit{b}) and 0.25~Myr
    (\textit{c} and \textit{d}).  Line profile parameters are
    calculated for each position-position pixel of a synthetic
    position-position-velocity cube of the \oiii{} line (\textit{a}
    and \textit{c}) and \nii{} line (\textit{b} and \textit{d}), which
    takes into account dust absorption (but not scattering) and
    thermal broadening at the local temperature. In each subfigure,
    results are shown for six viewing directions that correspond to
    the positive and negative principal axes of the simulation cube.
    The average thermal velocity width is subtracted in quadrature in
    order to mimic the observational methodology for isolating the
    non-thermal contribution to \(\sigma\los\). }
  \label{fig:simulation-vmean-sigma}
\end{figure}

\begin{figure}
  \centering
  \includegraphics[width=\linewidth]{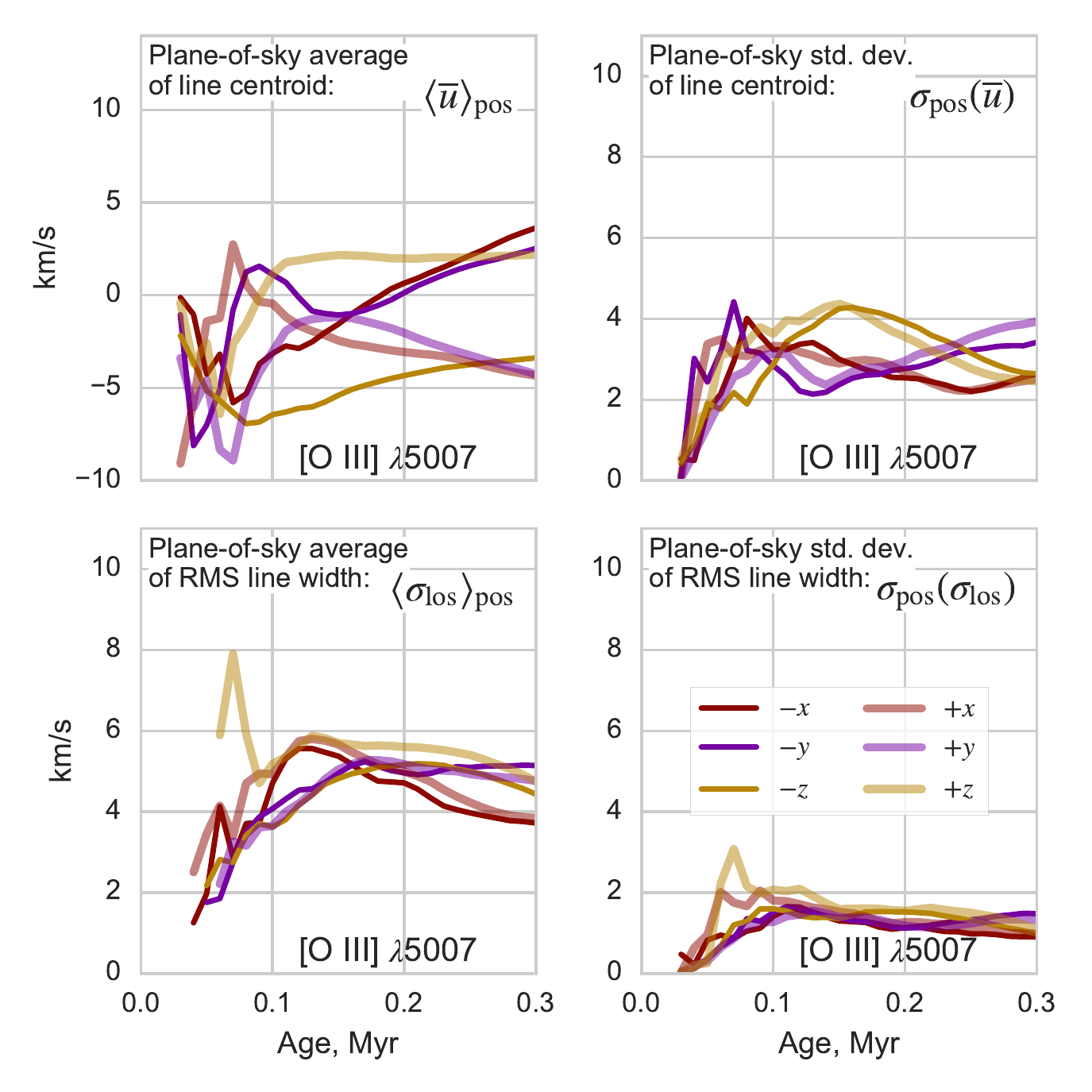}
  \caption{Temporal evolution of velocity statistics calculated from
    synthetic observed position-position-velocity cubes of the \oiii{}
    line, as illustrated in left panels of
    Fig.~\protect\ref{fig:simulation-vmean-sigma}.  Top row shows plane-of-sky
    average (left) and standard deviation (right) of the mean line
    velocity \(\bar{u}\), while bottom row shows the same for the
    non-thermal RMS line width \(\sigma\los\).  Different colored
    lines correspond to viewing directions along each of the cube
    axes.}
  \label{fig:simulation-stats-evo}
\end{figure}

For two representative times, Figure~\ref{fig:simulation-vmean-sigma}
shows the joint distribution of mean velocity \(\bar{u}\) and
non-thermal line width \(\sigma\los\), calculated from the line
profiles of synthetic position-position-velocity cubes for the
simulated turbulent \hii{} region of \citet{2014MNRAS.445.1797M}.
Figure~\ref{fig:simulation-stats-evo} shows the temporal evolution of
the plane-of-sky average and standard deviation of the same two
quantities for the \oiii{} line.  Results from the upper-right and
lower-left panel of Figure~\ref{fig:simulation-stats-evo} respectively
provide the data that go into the horizontal and vertical axes of
Figure~\ref{fig:obs-sigma-sigma}(\textit{b}) in
\S~\ref{sec:poslos}.

At late times (Fig.~\ref{fig:simulation-vmean-sigma}\textit{c,d}),
when dust absorption is relatively unimportant, the average line
centroid velocity (upper left panel of
Fig.~\ref{fig:simulation-stats-evo}) reflects the champagne flow due
to the largest-scale density gradients in our simulation box.  This
leads to both blue and red shifts, since opposite viewing directions
(e.g., \(+x\) and \(-x\)) have roughly equal but opposite mean
velocities, so that pairs of PDFs in
Fig.~\ref{fig:simulation-vmean-sigma}\textit{c,d} are rough mirror
images.  At earlier times
(Fig.~\ref{fig:simulation-vmean-sigma}\textit{a,b}), radial expansion
dominates and the dust optical depth is larger, which leads to
selectively greater absorption of receding regions of the nebula.
This produces predominantly blue-shifted mean velocities from all
viewing directions.

\section{Toy model of surface brightness profiles}
\label{sec:toy-model-surface}
\begin{figure}
  \centering
  \includegraphics[width=\linewidth]{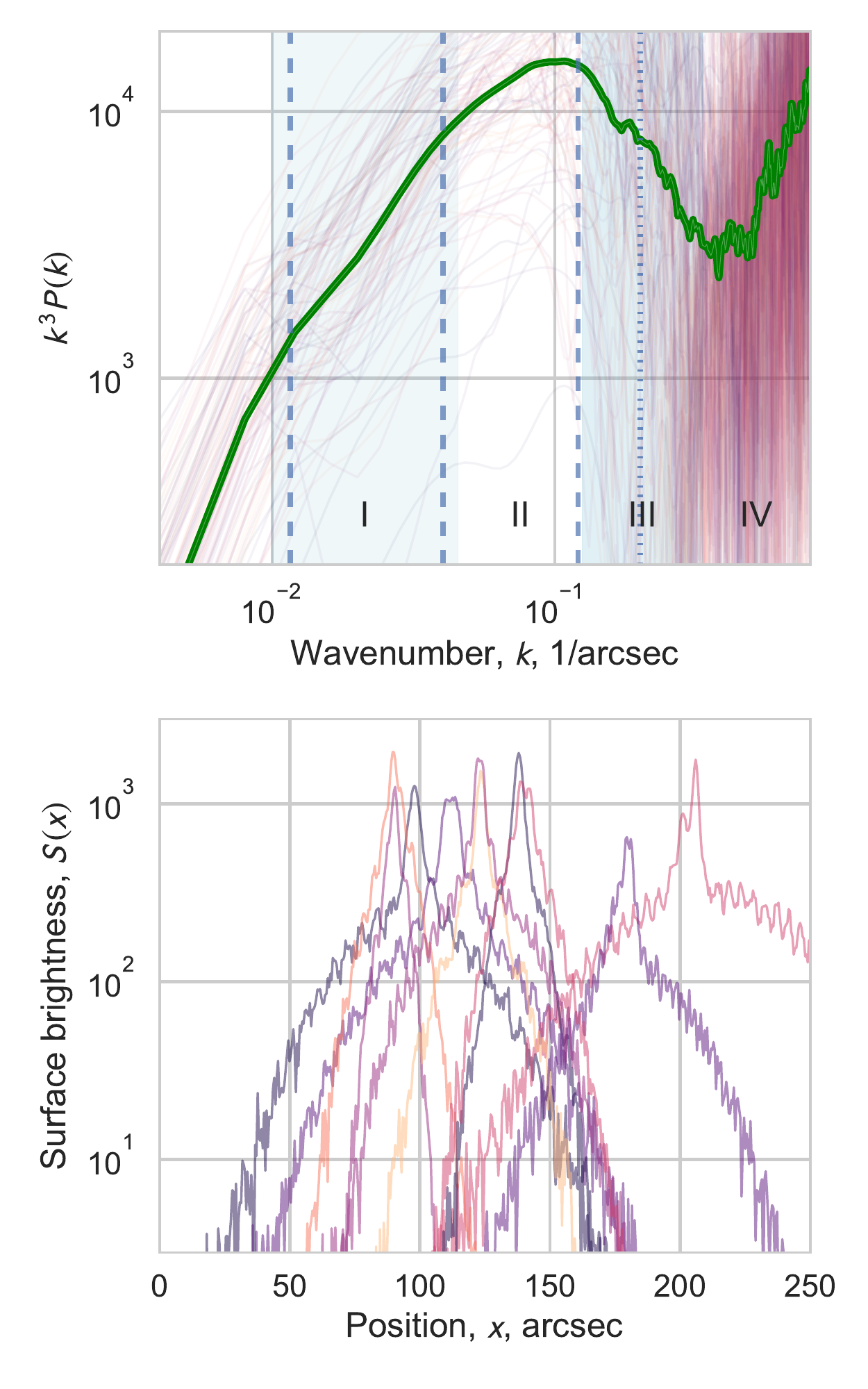}
  \caption{Compensated power spectra (upper panel) of 100 fake
    one-dimensional surface brightness profiles.  Individual power
    spectra are shown as faint lines, while the the thick green line
    shows the average spectrum.  The lower panel shows an example of
    10 of the brightness profiles on a semi-logarithmic scale.}
  \label{fig:fake-powerspec}
\end{figure}

To try and gain insight into the observed power spectra, we have
determined the simplest emission model that can reproduce the
qualitative features of regimes I--IV.  The emission model is
illustrated in Figure~\ref{fig:fake-powerspec} and consists of the
following spatial components:
\begin{enumerate}
\item A narrow Gaussian with peak brightness 800 (\(\pm 30\%\)) and FWHM
  \(4''\) (\(\pm 30\%\)).
\item A broader Gaussian with peak brightness 400 (\(\pm 30\%\)) and
  FWHM \(12''\) (\(\pm 30\%\)).
\item A very broad Gaussian with peak brightness 240 (\(\pm 30\%\)) and
  FWHM \(88''\) (\(\pm 50\%\)). 
\item A sinusoidal variation with wavelength \(5''\) (\(\pm
  50\%\)) and relative amplitude 15\%, which multiplies the sum of
  components (i)--(iii). 
\item Poisson \(\sqrt{N}\) noise, assuming that the brightness is in
  units of counts/pixel.
\end{enumerate}
The wavenumber \(k =
1/\lambda\) of each component is
indicated by vertical lines on the power spectrum shown in the upper
panel of Figure~\ref{fig:fake-powerspec} (dashed lines for the
Gaussian components (i)--(iii), dotted lines for the sinusoidal
component (iv)). For the Gaussian components, we assume \(\lambda
\approx 2 \times \mathrm{FWHM}\) since the Gaussian peak represents
only half of a fluctuation wavelength. 

Components (i) and (ii) are chosen so as to reproduce regime~II in the
power spectra of \S~\ref{subsec:vca}, with a relatively shallow slope
(\(\gamma > -3\)) between scales of 8\(''\) and 22\(''\).  As is, the
model best represents the \nii{} power spectrum (upper right panel of
Fig.~\ref{fig:HNOVCA}), but the precise value of the slope can be controlled by
varying the relative strength of the narrow (\(4''\)) and the broader
(\(12''\)) component.  For example, the \oiii{} power spectrum
requires nearly equal amplitudes for these two components.  

With only components (i) and (ii), the model power spectrum is too shallow
in regime~I (with \(\gamma \approx 0\)) and is too steep in
regime~III, showing a deep minimum in the \(k^3 P(k)\) curve before
arriving at the noise-dominated regime~IV.  These minor deficiencies are
ameliorated by the introduction of the secondary components~(iii)
and~(iv), respectively. 

The average power spectrum from 100 realizations of the toy model is
calculated, where the brightness and width of each component is chosen
according to normal distributions with central value (\(\pm\) RMS \%
variation) as given above.  The result is in remarkably good agreement
with the observed power spectra, except that the observations tend to
show sharper breaks at the boundaries between the different
regimes. It should be emphasized that the toy model is only an
idealization of the spatial variations present in the real surface
brightness maps, which typically show \(\sim 10\) prominent peaks in
each one-dimensional slit profile, rather than the single peak of the
model.

\label{lastpage}
\end{document}